%% file: CIB_bathtub.tex
\documentclass[a4paper,fleqn,usenatbib]{mnras} 
\usepackage{newtxtext,newtxmath}
\usepackage[T1]{fontenc}
\usepackage{ae,aecompl}

\include{mycommand}
\usepackage{mathtools,tabu}
\usepackage{amsmath,amssymb}

\title[Interpreting CFIRB using gas regulator]{Interpreting the cosmic far-infrared background anisotropies using a gas regulator model}
\author[Wu et al.]{Hao-Yi Wu,$^{1,2}$\thanks{Present address: The Ohio State University, Columbus, OH 43210, USA. E-mail: wu.3863@osu.edu}
Olivier Dor\'e,$^{1,2}\thanks{E-mail: olivier.p.dore@jpl.nasa.gov}$ 
Romain Teyssier$^{3}$
and Paolo Serra$^{1,2}$\\
$^{1}$California Institute of Technology, MC 367-17, Pasadena, CA 91125, USA \\
$^{2}$Jet Propulsion Laboratory, 4800 Oak Grove Drive, Pasadena, CA 91109, USA \\
$^{3}$Institute for Computational Science, University of Zurich, CH-8057 Z\"urich, Switzerland
}
\date{Accepted 2018 January 5. Received 2017 December 27; in original form 2016 July 7}
\pubyear{2017}
\begin{document}
\label{firstpage}
\pagerange{\pageref{firstpage}--\pageref{lastpage}}
\maketitle

\begin{abstract}
Cosmic far-infrared background (CFIRB) is a powerful probe of the history of star formation rate (SFR) and the connection between baryons and dark matter across cosmic time. In this work, we explore to which extent the CFIRB anisotropies can be reproduced by a simple physical framework for galaxy evolution, the gas regulator (bathtub) model. This model is based on continuity equations for gas, stars, and metals, taking into account cosmic gas accretion, star formation, and gas ejection. We model the large-scale galaxy bias and small-scale shot noise self-consistently, and we constrain our model using the CFIRB power spectra measured by {\em Planck}. Because of the simplicity of the physical model, the goodness of fit is limited. We compare our model predictions with the observed correlation between CFIRB and gravitational lensing, bolometric infrared luminosity functions, and submillimetre source counts. The strong clustering of CFIRB indicates a large galaxy bias, which corresponds to haloes of mass $10^{12.5}\Msun$ at $z=2$, higher than the mass associated with the peak of the star formation efficiency. We also find that the far-infrared luminosities of haloes above $10^{12}\Msun$ are higher than the expectation from the SFR observed in ultraviolet and optical surveys.
\end{abstract}


\section{Introduction}

Cosmic far-infrared background (CFIRB) originates from unresolved dusty star-forming galaxies across cosmic time. In these galaxies, the ultraviolet (UV) photons associated with newly formed, massive stars are absorbed by dust and re-emitted in far-infrared (FIR), and the FIR emission serves as an indicator of the star formation rate (SFR). At the FIR wavelengths ($\sim$100 \micron\ to 1\ mm, also known as submillimetre), most galaxies are unresolved and can only be observed as background intensity fluctuations. These fluctuations contain information about the cosmic star formation history, as well as the dark matter haloes in which the dusty star-forming galaxies are located. Compared with UV, the star formation history from FIR is much less explored because of the limited angular resolutions of the telescopes; thus, CFIRB provides an important piece of the puzzle of the cosmic star formation history.

Predicted half a century ago
\citep{PartridgePeebles67b,Bond86}, the CFIRB was first discovered by {\em COBE}-FIRAS \citep{Puget96,Fixsen98,Hauser98,Gispert00,HauserDwek01}
and subsequently observed by 
ISO \citep{LagachePuget00,Matsuhara00,Elbaz02}.
The anisotropies of CFIRB have been measured by
{\em Spitzer} \citep{GrossanSmoot07,Lagache07},
BLAST \citep{Viero09}, 
SPT \citep{Hall10}, 
ACT \citep{Hajian12}, 
{\em Herschel} \citep{Amblard11,Berta11,Viero13}, 
and {\em Planck} \citep{Planck11CIB, Planck13XXX}.
In particular, the angular power spectra of CFIRB provide the luminosity-weighted galaxy bias and thus the information about the mass of the underlying dark matter haloes \citep[e.g.,][]{Viero09,Amblard11,Planck11CIB,DeBernardis12,Shang12,Xia12,Thacker13,Viero13,Planck13XXX}.

To date, most of the interpretations of the CFIRB anisotropies are based on phenomenological models with limited physical interpretation. For example, \citet{Addison13} modelled the CFIRB and number counts using general parametrizations for the luminosity function, the spectral energy distribution (SED), and the scale-dependent galaxy bias. On the other hand, \citet{Shang12} implemented a luminosity--mass relation in the halo model to improve the modelling at small scales \citep[also see, e.g.,][]{Viero13}. In addition, \citet{Planck13XXX} provided updated measurements of the CFIRB power spectra as well as new constraints on linear and halo models; however, the SFR density inferred from their halo model appears higher at high redshift when compared with UV and optical observations.

In this work, we develop a physical model for the connection between dark matter haloes and dusty star-forming galaxies. We constrain this model using the CFIRB power spectra measured by {\em Planck}. We then compare our model with various FIR/submillimetre galaxy observations. Our model provides a simple, physically-motivated framework to compare and interpret various FIR observations. 

We apply the gas regulator model, which is based on the continuity equations of gas, stars, and metal \citep[also known as the bathtub or reservoir model, see, e.g.,][]{Bouche10,KrumholzDekel12,Dekel13,Lilly13,DekelMandelker14}, to calculate SFR. We then apply the halo model to calculate the power spectra of CFIRB \citep{ScherrerBertschinger91,Seljak00,CooraySheth02}. We fit the model to the CFIRB anisotropies measured by {\em Planck} \citep{Planck13XXX}. Our model predictions are compared with various IR observations, as well as the cosmic SFR density and cosmic dust mass density constrained by other observations. We find that CFIRB requires high IR luminosity for massive haloes ($\LIR \sim 10^{12}\Lsun$ for haloes of mass above $10^{13} \Msun$); this result is consistent with earlier findings \citep[e.g.,][]{Shang12,Addison13,Bethermin13} but is in excess compared with the SFR constrained by UV and optical. This excess of IR luminosity can be related to heating by old stellar populations.

This paper is organized as follows. 
Section~\ref{sec:bathtub} describes the gas regulator model and provides a quasi-steady-state solution relevant for SFR and dust property.
In Section~\ref{sec:halomodel}, we incorporate the gas regulator model into the halo model to calculate observed quantities. 
In Section~\ref{sec:fitting}, we fit our model to the CFIRB angular power spectra and intensity.
Section~\ref{sec:comparisons} shows comparisons between our model and other infrared observations.
In Section~\ref{sec:implications}, we discuss the implications of our model, including the galaxy--halo connection and the cosmic star formation history;
in Section~\ref{sec:discussion}, we discuss the limitations of our model and possible improvements.
We summarize in Section~\ref{sec:summary}.

Throughout this paper, we use a flat $\Lambda$CDM cosmology based on the {\em Planck} 2013 results \citep{Planck13cosmo};
$\Omega_M$ = 0.31;
$\Omega_\Lambda$ = 0.69;
$h$ = 0.67.
We use the linear matter power spectrum at $z=0$ calculated by {\sc CAMB} \citep{Lewis00} with 
$\Omega_b h^2 = 0.022$;
$\Omega_c h^2 = 0.12$;
$n_s = 0.96 $;
$A_s = 2.215\times10^{-9}$.
When converting SFR to IR luminosity, we use 
$\LIR = {\rm SFR}/{K}$, where $K=1.7\times10^{-10}\ \Msun\ \rm yr^{-1}\ \Lsun^{-1}$ based on the Salpeter initial mass function \citep{Kennicutt98}.

\section{Gas regulator model for galaxy evolution}\label{sec:bathtub}

In the gas regulator model, a galaxy is assumed to be a reservoir of gas, stars, and metal; the mass of each component is determined by a continuity equation with sources (cosmic accretion), sinks (star formation), and outflow. This model assumes that both the SFR and the gas outflow rate are proportional to the gas mass; therefore, the system is self-regulated and will eventually reach a steady state \citep[e.g.,][]{Bouche10,Dekel13,Lilly13}. Our model is based on the minimal implementation in \citet[DM14 thereafter]{DekelMandelker14} with various modifications. Table~\ref{tab:bathtub} summarizes the physical processes in this model, and Table~\ref{tab:parameters} lists the parameters in this model.

\subsection{Basic model and quasi-steady-state solution}

To describe the source terms, let us denote the cosmic accretion rate of {\em all} baryon mass as $\dMa$. In this accreted baryon mass, we assume that the gas mass fraction is $\fga$, and the stellar mass fraction is $(1-\fga)$. Star formation converts gas mass to stellar mass. We denote the SFR of the galaxy as $\dMsf$; since stars return a fraction (denoted as $R$) of the gas to the reservoir, the gas consumption rate is given by $(1-R)\dMsf$. In addition, the gas mass can be ejected from the galaxy due to feedback processes, and we assume that the mass-loss rate is proportional to the SFR, $\eta \dMsf$. Here, $\eta$ is the mass-loading factor and will be discussed in detail in Section~\ref{sec:massloading}. We assume that the outflow of stellar mass is negligible.

The continuity equations of gas mass ($\Mg$) and stellar mass ($\Ms$) are given by
\beq
\dMg = f_{\rm ga} \dMa - (1-R + \eta ) \dMsf \ ,
\label{eq:dMg}
\eeq
and 
\beq
\dMs = (1-f_{\rm ga})\dMa + (1-R)\dMsf \ .
\eeq
Since the stellar mass is not directly observable in FIR, we will not further discuss the stellar mass in this paper.

We assume that the cosmic accretion provides negligible metal mass. The metal production rate is given by $y(1-R)\dMsf$, where $y$ is the metal yield\footnote{In this work, we define the metal yield $y$ as the ratio between the metal mass returned to the gas and the stellar mass locked in stars \citep[e.g.,][]{Schneider}.}. The loss of metal is proportional to the loss of gas. The continuity equation of metal mass ($\Mm$) is thus given by
\beq
\dMm = y (1-R) \dMsf - (1 -R + \eta)\dMsf \frac{\Mm}{\Mg} \ .
\label{eq:dMm}
\eeq

For the quasi-steady-state solution, we assume 
$\dMg = 0$ and $\dMm = 0$. Equations (\ref{eq:dMg}) and (\ref{eq:dMm}) become
\beq
\dMsf = \frac{\fga\dMa}{1-R+\eta} \label{eq:SFR} 
\eeq
and 
\beq
\Mm = \Mg \frac{y(1-R)}{1-R+\eta} \ .
\label{eq:Mm}
\eeq
Under this assumption, the gas metallicity $\Mm/\Mg$ is constant with time.

To calculate the gas mass, we assume that $\dMsf = \Mg/\tsf$, where $\tsf$ is the star formation time-scale, 
\beq
\Mg = \frac{\fga\dMa \tsf}{1-R+\eta} \ .
\eeq

\begin{table*}
\hspace{-1.5cm}
\setlength{\tabcolsep}{0.5em}
\begin{tabular}{c||c|c|c}
\hline\hline
\rule[-2mm]{0mm}{6mm} Physical process & Gas & Star & Metal in gas \\ \hline
\rule[-2mm]{0mm}{6mm} Cosmic accretion & $\fga\dMa$ & $(1-\fga)\dMa$ & (Negligible) \\ 
\rule[-2mm]{0mm}{6mm} Star formation & $-(1-R)\dMsf$& $(1-R)\dMsf$ & $y(1-R)\dMsf -(1-R)\dMsf \Mm/\Mg$\\ 
\rule[-2mm]{0mm}{6mm} Outflow & $-\eta\dMsf$ & (Negligible) & $-\eta \dMsf \Mm/\Mg$ \\ \hline
\hline
\end{tabular}
\caption{Summary of source, sink, and outflow terms in the gas regulator model.}
\label{tab:bathtub}
\end{table*}

\begin{table*}
\vspace{1cm}
\setlength{\tabcolsep}{0.5em}
\begin{tabular}{clcl}
\hline\hline
\rule[-2mm]{0mm}{6mm} Parameter & Meaning & Fiducial value & Reference \\
\hline\hline
\rule[-2mm]{0mm}{6mm} & Cosmic accretion &&\\ \hline
\rule[-2mm]{0mm}{6mm} $f_b$ & $\Omega_b/\Omega_M$ & 0.18 & \citet{Planck13cosmo} \\
\rule[-2mm]{0mm}{6mm} $\fga$ & (gas mass) / (gas mass + stellar mass) in cosmic accretion, $0<\fga < 1$ & 0.8 & \citet{DekelMandelker14} \\
\rule[-2mm]{0mm}{6mm} $\dMa$ & Accretion rate of {\em all} baryon mass & -- & ibid. \\
\rule[-2mm]{0mm}{6mm} $p$ & Penetration factor, $M_{\rm accreted\ baryon}/(f_b M_{\rm accreted\ DM})$ & 0.5 & ibid. \\
\hline\hline
\rule[-2mm]{0mm}{6mm} &Star formation &&\\ \hline
\rule[-2mm]{0mm}{6mm} $K$ & $\LIR = {\rm SFR}/K$ & $1.7\times10^{-10}$ & \citet{Kennicutt98} \\ 
\rule[-2mm]{0mm}{6mm} $\dMsf$ & SFR & -- & \citet{DekelMandelker14}\\ 
\rule[-2mm]{0mm}{6mm} $\tsf$ & Star formation time-scale $\tsf = \epsilon^{-1} t_d$ & -- & ibid. \\ 
\rule[-2mm]{0mm}{6mm} $\epsilon$ & SFR efficiency per dynamical time & 0.02 & ibid. \\ 
\rule[-2mm]{0mm}{6mm} $t_d$ & Dynamical time, $t_d = \nu_d t$, where $t$ is the cosmic time & -- & ibid. \\ 
\rule[-2mm]{0mm}{6mm} $\nu_d$ & $t_d$ in units of the cosmological time & 0.0071 & ibid. \\ 
\rule[-2mm]{0mm}{6mm} $R$ & Fraction of gas mass returned by star formation & 0.46 & ibid. \\ 
\rule[-2mm]{0mm}{6mm} $\eta$ & Mass-loading factor, ratio between gas outflow and SFR & -- & equation (\ref{eq:etaM}) \\ 
\hline\hline
\rule[-2mm]{0mm}{6mm} & Metal and dust &&\\ \hline
\rule[-2mm]{0mm}{6mm} $y$ & Metal yield & 0.016 & \citet{Lilly13} \\%
\rule[-2mm]{0mm}{6mm} $r$ & Dust-to-metal mass density ratio & 0.4 & \citet{Hayward11} \\ 
\hline\hline
\rule[-2mm]{0mm}{6mm} & Dust SED &&\\ \hline
\rule[-2mm]{0mm}{6mm} $\beta$ & Spectral index of dust SED & (2) & ibid. \\ 
\rule[-2mm]{0mm}{6mm} $\kappa$ & Dust opacity, $\kappa = \kappa_0 (\nu/\nu_0)^\beta$ & -- & \citet{Hayward11}\\
\rule[-2mm]{0mm}{6mm} $\kappa_0$ & Opacity at the pivot frequency & 0.050 & ibid. \\
\rule[-2mm]{0mm}{6mm} $\nu_0$ & Pivot frequency for opacity & 850$\micron$ & ibid. \\ 
\hline\hline
\rule[-2mm]{0mm}{6mm} &Halo mass -- IR luminosity relation &&\\ \hline 
\rule[-2mm]{0mm}{6mm}$\Mpk$ & Peak halo mass for SFR & $10^{12}\Msun$ & \cite{Behroozi13} \\ 
\rule[-2mm]{0mm}{6mm}$\Mmin$ & Minimum halo mass for hosting a FIR galaxy & $10^{11}\Msun$ & \cite{KrumholzDekel12} \\
\hline\hline
\end{tabular}
\caption{Parameters in the gas regulator model. Numbers in parentheses indicate the values used in the references; these parameters are set free in our model.}
\label{tab:parameters}
\end{table*}

\subsection{Implementation}
Equation (\ref{eq:SFR}) is our prediction for the SFR.
We assume that the baryon mass accretion rate $\dMa$ is proportional to the dark matter accretion rate
\beq
\dMa = f_b p \dMh \ ,
\eeq
where $\Mh$ is the mass of the dark matter halo; 
$f_ b$ is the cosmic baryon mass fraction $\Omega_b / \Omega_M$, which is assumed to be 0.18 \citep{Planck13cosmo};
$p$ indicates the mass fraction of the gas that can penetrate the halo and reach
the galaxy.

For the mass accretion rate of dark matter haloes, we use the 
fitting formula calibrated using the two Millennium simulations
by \citet{Fakhouri10}
\beqa
\dot{\Mh} = &46.1 {\Msun yr^{-1}} \bigg( \frac{M}{10^{12}\Msun} \bigg)^{1.1}\\ 
&\times (1+1.11 z )\sqrt{\Omega_M(1+z)^3 + \Omega_\Lambda} .
\label{eq:MAH}
\eeqa
We include an extra redshift dependence to model the fact that the SFR does not necessarily trace the gas accretion rate,
\beq\label{eq:fz}
f(z) = (1+z)^\delta \ .
\eeq
We assume that $\dMsf$ is proportional to the IR luminosity, 
\beq
\LIR = \frac{\dMsf}{K} \ , 
\eeq
where $K=1.7\times10^{-10}\ \Msun\ \rm yr^{-1}\ \Lsun^{-1}$ \citep[][based on the Salpeter initial mass function]{Kennicutt98}.

To summarize, the $\LIR$--halo mass relation is given by
\beq
\LIR = \frac{\fga f_b p}{K(1-R+\eta)} \dot{\Mh} f(z) 
\label{eq:LIR} .
\eeq

We assume that the dust mass is proportional to the metal mass with a constant dust-to-metal ratio,
$r= \Md/\Mm$, and is given by
\beq
\Md = \frac{r y (1-R)}{1-R+\eta} \tsf \dMsf = \frac{r y (1-R)}{(1-R+\eta)^2} \fga f_b p \dot{\Mh} f(z) \tsf \ .
\label{eq:Mdust}
\eeq
Following DM14, we assume that the star formation time-scale is proportional to the dynamical time, 
$\tsf = \epsilon^{-1} t_d $, and $\epsilon = 0.02$. The dynamical time is assumed to be proportional to the cosmic time, $t_d = \nu t$, and $\nu= 0.0071$.

We assume that the spectral luminosity is given by an optically-thin modified blackbody with a single dust temperature $\Td$ \citep[e.g.,][]{Hayward11} 
\beq
L_\nu = 4{\rm\pi}\kappa_\nu \Md B_\nu(\Td) \ ,
\label{eq:Lnu}
\eeq
and that the opacity in IR follows a power-law
\beq
\kappa_\nu = \kappa_0 \bigg( \frac{\nu}{\nu_0} \bigg)^\beta \ .
\label{eq:opa}
\eeq
Integrating $L_\nu$ over $\nu$, we obtain $\LIR$ as a function of $\Md$ and $\Td$.
Solving for $\Td$, we obtain 
\beq
\Td = \frac{h}{k}\bigg[\frac{\LIR c^2 \nu_0^\beta}
{\Gamma(4+\beta) \zeta(4+\beta) 8{\rm\pi} \kappa_0 h \Md}
\bigg]^{1/(4+\beta)} \ .
\label{eq:Tdust}
\eeq
Following \citet{Hayward11}, we assume that $\kappa_0 = 0.07\ {\rm m^2 kg^{-1}}$ at $850\ \micron$ at observed frame, $\nu_0 = 353\ (1+z)\ {\rm GHz}$. The spectral index $\beta$ is a free parameter in our model.
Since we are only concerned with the FIR wavelengths in the Rayleigh--Jeans tail, we expect that 
the single-temperature modified blackbody is a reasonable description for our SED.

\subsection{Modelling feedback via mass-loading factor}\label{sec:massloading}

Equation (\ref{eq:SFR}) indicates that the SFR is determined by the mass accretion rate; however, additional feedback processes can affect the SFR. For low-mass haloes, supernova feedback can eject gas efficiently and suppress the SFR \citep[e.g.,][]{Benson03,DuttonvdBosch09}. To model this effect, we assume $\eta \propto \Mh^{-\alpha_1}$ for $\Mh < \Mpk$, where $\Mpk$ is the halo mass associated with the peak of the star formation efficiency.

Different values of $\alpha_1$ correspond to different physical models for supernova feedback. For example, for energy-driven winds, $\eta\propto \Vvir^{-2} \propto \Mvir^{-2/3}$ \citep[e.g.,][]{Benson10}; for momentum-driven winds, $\eta\propto \Vvir^{-1} \propto \Mvir^{-1/3}$ \citep[e.g.,][]{Murray05,Hopkins12}; for constant winds, $\eta$ = constant \citep[e.g.,][]{SpringelHernquist03}. Steeper scaling relations have also been adopted by some semi-analytical models \citep[e.g.,][]{Guo11}. Observations have been used to estimate the velocities of gas outflow; however, constraining the mass dependence of the mass-loading factor is still challenging \citep[e.g.,][see \citealt{Veilleux05,Erb15} for reviews]{Weiner09,Chen10,Martin12,Rubin14}. 

For massive haloes, the SFR is suppressed by feedback from active galactic nuclei \citep[e.g.,][]{Croton06} or quenched due to environment \citep[e.g.,][]{Wetzel12}. Thus, for massive haloes ($\Mh > \Mpk$), we phenomenologically model the mass-loading factor as $\eta\propto M^{\alpha_2}$; this parametrization effectively describes the reduced supply of cold gas. In addition, observations have hinted that SFR and the AGN luminosity is related to each other \citep{Lutz10}, supporting the gas regulator model in the regime of AGN feedback.

To make the transition between high- and low-mass smooth, we adopt the function form \citep[see, e.g.,][]{Feldmann15}
\beq
\eta(M) = f(x,y) = \eta_0 \bigg(1+x+y-(1+x^{-1}+y^{-1} )^{-1} \bigg) \ , 
\label{eq:etaM}
\eeq
where 
\beq
x = \bigg( \frac{\Mh}{M_{\rm pk}}\bigg)^{-\alpha_1} ,\quad
y = \bigg( \frac{\Mh}{M_{\rm pk}}\bigg)^{\alpha_2} \ .
\eeq
We assume $\Mpk = 10^{12}\Msun$ (see e.g., \citealt{Behroozi13}) and use 
three free parameters to describe the mass-loading factor: ($\eta_0$, $\alpha_1$, $\alpha_2$).

\section{Halo model for clustering}\label{sec:halomodel}
Given the $\LIR$--$\Mh$ relation and the SED from the gas regulator model, we can apply the halo model to calculate the CFIRB power spectra and various FIR observables. We include the scatter between IR luminosity and halo mass\footnote{We note that in the presence of a scatter, all equations in Section~\ref{sec:halomodel-CL} involve $\int L P(\lnL|M) {\rm d}\lnL = \avg{L}$; therefore, all the equations in this section look the same as if there is no scatter.}.

\subsection{CFIRB Intensity and power spectra}\label{sec:halomodel-CL}

We denote $\nu$ as the frequency in the {\em observed} frame. For brevity, we denote $\LIR$ as $L$ and $\Mh$ as $M$ below. The emission coefficient at $\nu$ at redshift $z$ is given by integrating the spectral luminosity of all haloes, described by the halo mass function ($dn/dM$), at this redshift,
\beq
j_\nu(z) = \int dM \frac{dn}{dM} f_\nu(M,z) \ ,
\label{eq:jz}
\eeq
where
\beq
f_\nu(M,z) = \frac{1}{4{\rm\pi}} \Lnuobs(M,z) \ .
\eeq
We note that here $\Lnuobs$ includes the contribution from both central and satellite galaxies, because in the gas regulator model we calculate the accretion rate of the entire host halo. This is a major difference between our model and the model in \cite{Shang12}.

The spectral intensity is given by integrating the emission coefficient over all redshifts,
\beq
I_{\nu} = \int dz \frac{d\chi}{dz} a j(z) \ ,
\eeq
where $a = 1/(1+z)$ is the scale factor, and $\chi$ is the comoving distance.

The angular power spectra at large scale are determined by galaxy pairs in two different haloes, i.e., 
the two halo term, which is given by
\beq
C^{2h}_{\ell,\nu{\nu'}} = \int \frac{dz}{\chi^2} \frac{d\chi}{dz} a^2 B_\nu(z)B_{\nu'}(z) P_{\rm lin}{\left(k=\ell/\chi,z\right)} \ ,
\eeq
where $B_\nu$ is given by 
\beq
B_\nu(z) = \int dM \frac{dn}{dM} b(M) f_\nu(M,z) \ ,
\label{eq:Bz}
\eeq
where $b(M)$ is the halo bias; we use the fitting function in \cite{Tinker10}.

The contribution by galaxy pairs in the same halo, i.e., the 1-halo term, is given by
\beq
C^{1h}_{\ell,\nu{\nu'}}= \int \frac{dz}{\chi^2} \frac{d\chi}{dz} a^2 A_{\nu\nu'}(k,z) \ ,
\eeq
where
\beq
A_{\nu\nu'}(k,z) = \int dM \frac{dn}{dM} f_\nu f_{\nu'} u^2 \ . 
\label{eq:Az}
\eeq
Here $u = u(k,M,z)$ is the halo mass density profile in Fourier space; we adopt the NFW profile \citep{NFW97}.

\subsection{Spectral flux density function and shot noise}\label{sec:fluxfunc}

The spectral flux density is related to the spectral luminosity via
\beq 
S_\nu = \frac{L_{(1+z)\nu}}{4{\rm\pi} \chi^2 (1+z)} \ .
\eeq
We assume that at a given halo mass $M$, $S_\nu$ has the following probability distribution function
\beq
P(\lnSnu |M) = \frac{1}{\sqrt{2{\rm\pi}}\sigma}\exp\bigg[ -\frac{\big(\lnSnu - \avg{\lnSnu}\big)^2}{2\sigma^2}
\bigg] \ .
\label{eq:pdf}
\eeq
We note that under this assumption
\beq
\avg{\lnSnu} = \ln \avg{S_\nu} -\frac{\sigma^2}{2} \ .
\eeq
As we will see later, since $\sigma$ is not negligible, 
$\avg{\lnSnu} \neq {\rm ln}\avg{S_\nu}$.

The flux density function is given by integrating over the halo mass function 
\beqa
\frac{dn}{d\lnSnu}(z) &= \int dM \frac{dn}{dM} P(\lnSnu | \lnM) \ .
\eeqa

The shot noise of the power spectra is calculated by integrating the square of the flux density for all galaxies,
\beqa\label{eq:shot}
C^{\rm shot}_{\nu\nu} = \int dz\frac{d\chi}{dz}\chi^2 \int d\lnSnu \frac{dn}{d\lnSnu} S_{\nu} S_{\nu} \ .
\eeqa
For the shot noise in cross power spectra ($\nu\neq\nu'$), we assume
\beq
C^{\rm shot}_{\nu\nu'} = \bigg(C^{\rm shot}_{\nu\nu} C^{\rm shot}_{\nu'\nu'} \bigg)^{1/2} .
\eeq
This assumption is consistent with the cross shot noise found in \citet{Planck13XXX}.
We do not take into account the decorrelation between different frequencies, and this decorrelation has
 been constrained to be less than 1 per cent \citep{Mak17}.

\section{Fitting Model to CFIRB}\label{sec:fitting}

We present the data sets we use, our fitting procedure, and the constraints on model parameters.

\subsection{Observed CFIRB power spectra and intensity}

We use the angular power spectra published in \citet{Planck13XXX}, which are based on maps measured in four frequency bands by {\em Planck} High Frequency Instrument (HFI): 217, 353, 545, and 857 GHz (1382, 849, 550, and 350 $\micron$), for a total area of 2240 deg$^2$. In particular, we used the 10 auto- and cross- spectra presented in table D.2 in \citet{Planck13XXX}, which exclude the primordial cosmic microwave background (CMB), Galactic dust, and the thermal Sunyaev--Zeldovich effect. We use the multipoles $187\leq \ell \leq 2649$; this leads to 83 data points in total. We use the colour-correction factors given in Section 5.3 of \citet{Planck13XXX}.

\begin{figure}
\includegraphics[width=1\columnwidth]{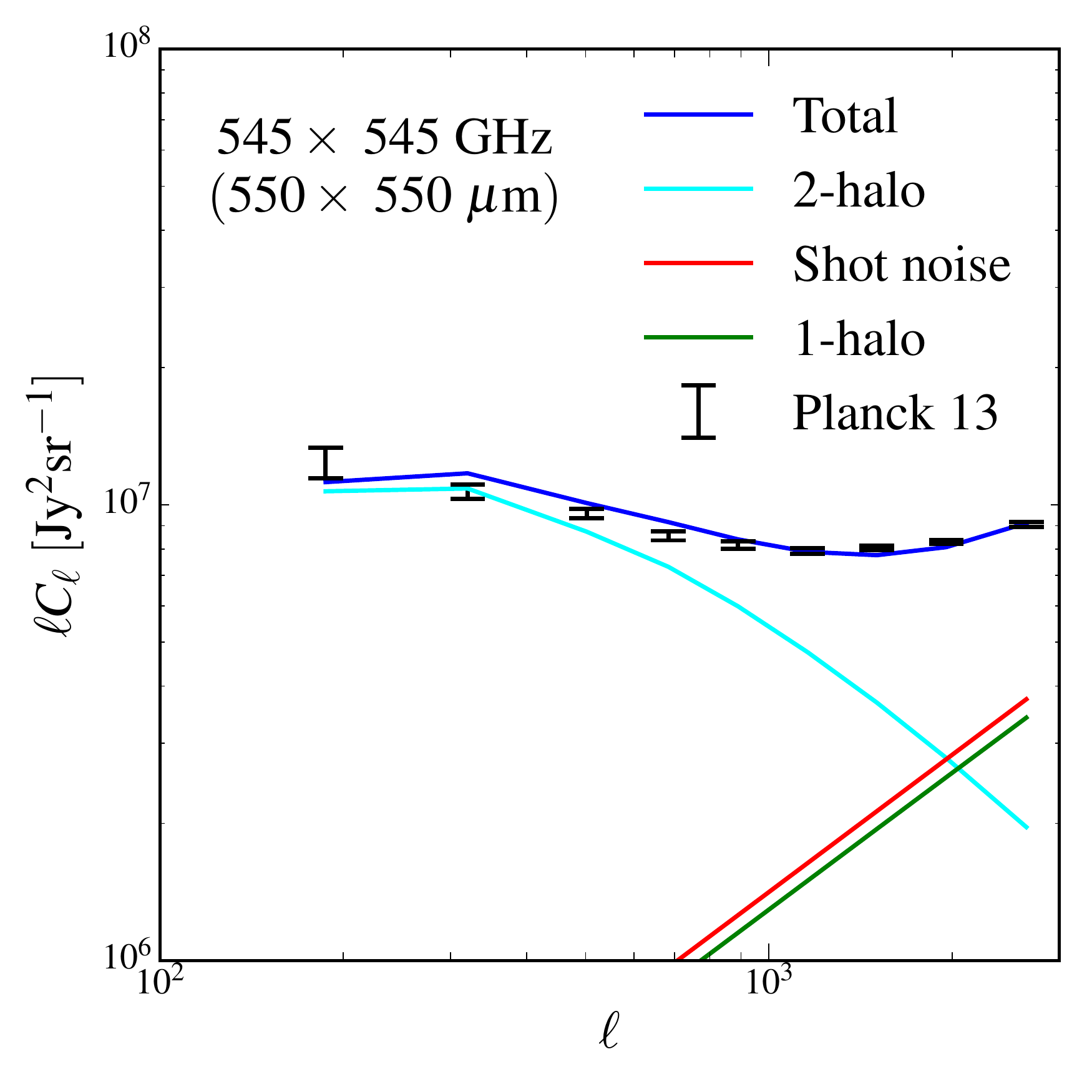}
\caption[CFIRB power spectra]{Model fit to the CFIRB angular power spectra measured by \protect\citet{Planck13XXX}. The blue curve is our model, which is broken down into the 2-halo term (cyan), the shot noise (red), and the 1-halo term (green). See Fig.~\ref{fig:CL_full} for the auto and cross power spectra for four frequency bands.}
\label{fig:CL}
\end{figure}
\begin{figure}
\includegraphics[width=1\columnwidth]{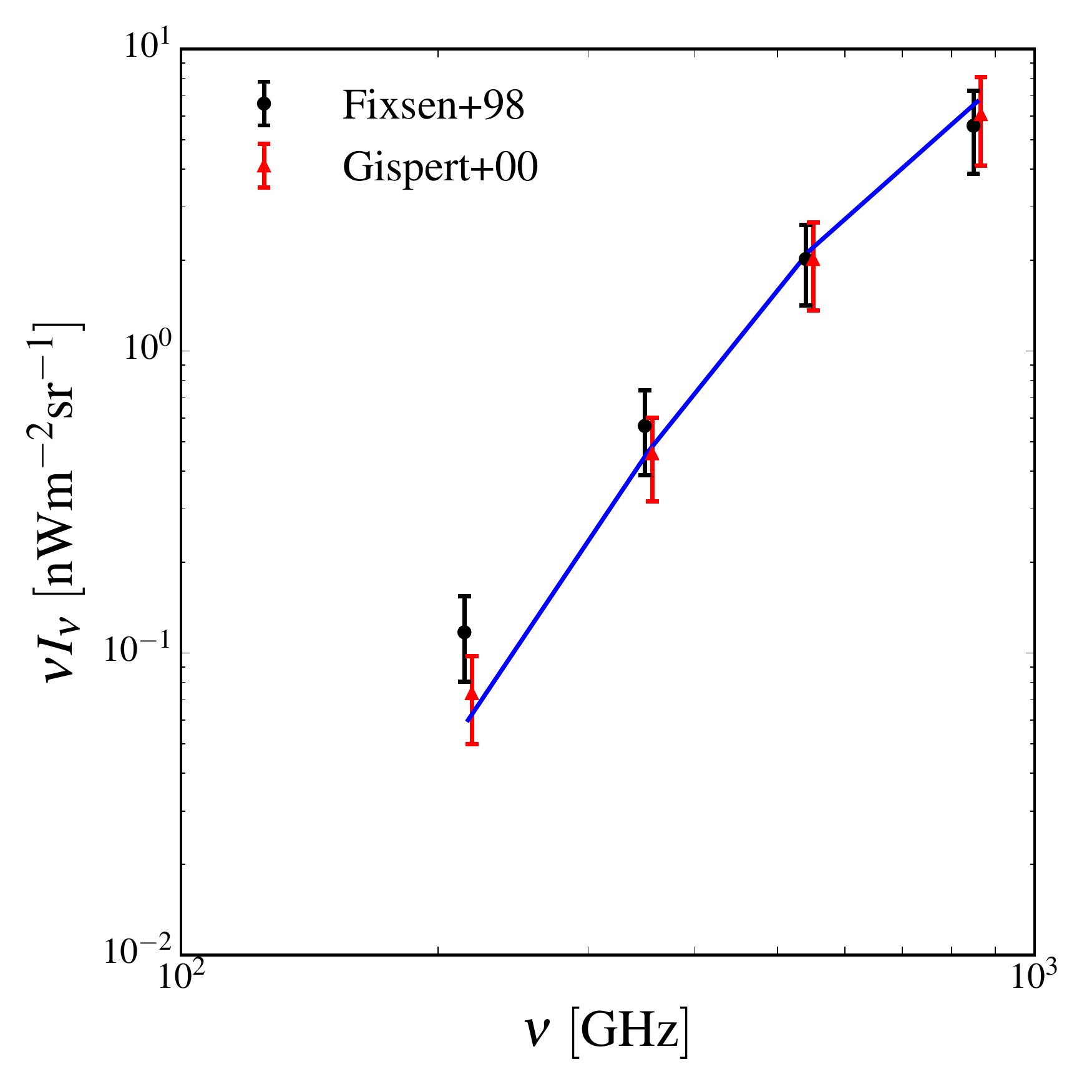}
\caption[CFIRB intensity.]{Prediction of the CFIRB intensity from our model compared with the measurement of {\em COBE}-FIRAS \protect\citep{Fixsen98,Gispert00}. We use the four frequencies associated with the {\em Planck}-HFI bands. The blue curve presents our best-fitting model.}
\label{fig:Inu}
\end{figure}

\subsection{Fitting procedure}
Our likelihood function $P({\bf D}|\theta)$ is given by
\beq
-\ln P({\bf D}|\theta) = \sum \frac{(D_i - M_i(\theta))^2}{\sigma_i^2} \ ,
\eeq
where $D_i$ is a data point, $\sigma_i$ is its error bar, and $M_i$ is the model prediction based on a set of parameters $\theta$. For the CFIRB angular power spectra, $D_i$ is $C_\ell^{\nu\nu'}$ and $\sigma_i$ is $\sigma(C_\ell^{\nu\nu'})$ for four auto- and six cross- spectra, for $\ell$ between 187 and 2649. 

We use the publicly available Markov chain Monte Carlo (MCMC) code {\sc emcee} \citep{emcee} version 2.0.0 to explore the parameter space. In particular, {\sc emcee} uses an ensemble of $N$ walkers to update each other. Briefly, for a given walker at position $X_k$, the algorithm uses another walker $X_{j\neq k}$ to propose a new position $Y = X_j + Z[X_k - X_j]$, where $Z$ is a random variable drawn from a distribution function that makes the proposal symmetric. The new position is accepted with a probability of $\min\left(1, Z^{N-1}p(Y)/p(X_k)\right)$, where $p(x)$ is the posterior probability. We refer the readers to \citet{emcee} for the complete description of the algorithm. 

We have six free parameters in the gas regulator model (see Table~\ref{tab:constraints}), 87 data points, and the $\chi^2$ is $\chisqBest$ for $\dof$ degrees of freedom. We use top-hat priors with generous ranges for all parameters. We have run 10 MCMC chains, each of which includes approximately 200,000 samples. We discard the first half of the chains as burn-in. We then apply the Gelman--Rubin diagnostic \citep{GelmanRubin92}, which compares the ``within-chain variance'' and the ``between-chain variance'' for multiple chains. We have ensured that the scale reduction factor $\sqrt{\hat R}$ is much less than 1.1. Table~\ref{tab:constraints} shows the constraints on the model parameters, and Table~\ref{tab:cor} shows the correlation matrix for these parameters. Fig.~\ref{fig:MCMC} shows the posterior distributions from the MCMC chains. 

Our best-fitting $\chi^2$ is larger than that in \cite{Planck13XXX}, which is 100.7 for 98 degrees of freedom,
including the 3000 GHz data and using free parameters to model the shot noise. Here we model the shot noise self-consistently but was unable to achieve such small $\chi^2$; therefore, our model should be regarded as qualitative rather than quantitative.

\begin{table*}
\setlength{\tabcolsep}{0.5em}
\begin{tabular}{ccclc}
\hline\hline
\rule[-2mm]{0mm}{6mm} Parameter & Prior & Constraint (68\%) & Definition & Equation \\ \hline\hline
\rule[-2mm]{0mm}{6mm}	$\eta_0$ & $\etaPri$ & $\etaCon$& Minimum value of mass-loading factor (at $\Mpk$) & \ref{eq:etaM} \\
\rule[-2mm]{0mm}{6mm}	$\alpha_1$ & $\alphaOnePri$ & $\alphaOneCon$& Slope of mass-loading factor for low-mass end ($\eta\propto M^{-\alpha_1}$) & \ref{eq:etaM} \\
\rule[-2mm]{0mm}{6mm}	$\alpha_2$ & $\alphaTwoPri$& $\alphaTwoCon$& Slope of mass-loading factor for high-mass end ($\eta\propto M^{\alpha_2}$) & \ref{eq:etaM} \\
\rule[-2mm]{0mm}{6mm}	$\beta$ & $\betaPri$ & $\betaCon$& Spectral index for dust opacity & \ref{eq:opa} \\
\rule[-2mm]{0mm}{6mm}	$\sigma$ & $\sigmaPri$ & $\sigmaCon$& Logarithmic scatter of $\LIR$ at a given halo mass & \ref{eq:pdf} \\
\rule[-2mm]{0mm}{6mm}	$\delta$ & $\deltaPri$ & $\deltaCon$& Extra redshift dependence of accretion rate & \ref{eq:fz} \\
\hline\hline
\end{tabular}
\caption{Constraints on the model parameters.}
\label{tab:constraints}
\end{table*}

\subsection{Best-fitting model}

Fig.~\ref{fig:CL} shows the data and the best-fitting model (with the maximum likelihood) for the CFIRB 
 the angular power spectrum at 545 GHz (550 $\micron$). Fig.~\ref{fig:CL_full} shows the full 10 auto- and cross-spectra from the four bands of {\em Planck}. We demonstrate the contribution from the 2-halo term, 1-halo term, and the shot noise. For the angular scale measured by {\em Planck}, the 1-halo term is sub-dominant. In Fig.~\ref{fig:CL_full}, we can see that the agreement is good for almost all angular scales and all bands. The fit for the 217 GHz (1382 $\micron$) auto-power spectrum is noticeably worse than other frequencies, which could be caused by our simplistic assumption of SED. We note that this band is dominated by CMB at all scales, and that the power spectrum can be affected by the procedure used for removing CMB.

Fig.~\ref{fig:Inu} shows the CFIRB intensity calculated from the best-fitting model. We show the data from both \citet{Fixsen98} and \citet{Gispert00}, using the values and error bars quoted in table 5 in \citet{Planck11CIB}. Our best-fitting model agrees better with \citet{Gispert00}, and we note that the result from \citet[][see their fig.\ 15]{Planck11CIB} also agrees better with \citet{Gispert00}.

\subsection{Constraints on model parameters}

In the following we discuss the implications of the constraints for our model parameters. We quote the median and the 68\% constraints for the 1-D marginalized posterior distribution. Table~\ref{tab:constraints} lists the parameter constraints.

\begin{itemize} 

\item $\eta_0$ (minimum of the mass-loading factor, which occurs at $\Mpk$). The constraint is $\etaCon$. As mentioned in Section~\ref{sec:massloading}, several observations provided a lower limit for the mass-loading factor but the observed values are inconclusive.

\item $\alpha_1$ (the slope of the mass-loading factor at low-mass end): The constraint is $\alphaOneCon$, which implies $\eta \propto \Mh^{-\alphaOneBest} \propto \Vvir^{-\alphaOneBestTimesThree}$. This scaling is much steeper 
compared with any of the supernova wind models. Our model prefers a very low $\LIR$ for low-mass haloes, which can be related to low SFR and/or low dust content. It has been shown that low-mass haloes tend to have a $\LIR$ lower than expected from the SFR due to the low mass content \citep[e.g.,][]{Hayward14}. 

\item $\alpha_2$ (the slope of the mass-loading factor at high-mass end): The constraint is $\alphaTwoCon$. As it is less than 1.1, the SFR does not decrease at the high-mass end (see equation \ref{eq:SFR} and Fig.~\ref{fig:LM}). We will further discuss this trend in Section~\ref{sec:LM}.

\item $\beta$ (slope of opacity, emissivity index): The constraint is $\betaCon$, which is close to the value $\beta = 2$ expected from theory \citep{DraineLee84}. It is higher than the results in \citet[][$\beta$=1.75]{Planck13XXX} and the nearby late-type galaxies observed by {\em Herschel} in \citet[][$\beta$=1.5]{Boselli12}.

\item $\sigma$ (scatter of $\lnSnu$ and $\ln\LIR$ at a given halo mass): The constraint is $\sigmaCon$ ($\sigmaDex$ dex). This parameter is constrained by the shot noise; as we will see later, it also reproduces the bright-end of the IR luminosity functions (Fig.~\ref{fig:LF}). We note that this scatter is smaller than our current knowledge of SFR. For example, the scatter between stellar mass and halo mass is estimated to be 0.2 dex \citep[e.g.,][]{Reddick13}, and the scatter between SFR and stellar mass is estimated to be 0.15 dex \citep[e.g.,][]{Bernhard14}; summing these two scatter values in quadrature will lead to a scatter of 0.25 dex between SFR and halo mass.

\item $\delta$ (extra redshift dependence, equation \ref{eq:fz}): The constraints is $\deltaCon$.
This value deviates from zero, indicating that the dark matter accretion rate (equation \ref{eq:MAH}) is insufficient to account for the full evolution of the SFR--mass relation. 
Our overall redshift dependence is approximately $(1+z)^{3.6}$ (see equation \ref{eq:gz} below), which is consistent with the results of \cite{Planck13XXX}.

\end{itemize}

\subsection{Summary of our model}

Here we summarize the main scaling relations based on our parameter constraints.
The $\LIR$--mass relations is given by
\beq
\label{eq:LM_final}
\LIR(M, z)= \frac{\LIRnorm}{1+\factorfM f(M)} \left(\frac{M}{10^{12}}\right)^{1.1} g(z) \ \Lsun \ .
\eeq
The dust mass is given by
\beq
\label{eq:Mdust_final}
\Md (M,z)= \frac{\Mdnorm}{\left(1+\factorfM f(M)\right)^2} \left(\frac{M}{10^{12}}\right)^{1.1} \left(\frac{t}{{\rm Gyr}} \right) g(z)\ \Msun \ ,
\eeq
and the dust temperature is given by
\beq
\label{eq:Tdust_final}
\Td(M,z) = \Tdnorm \left(\frac{1+\factorfM f(M)}{t/{\rm Gyr}}\right)^{1/(4+\beta)} \ {\rm K} \ .
\eeq
In the equations above, the extra time dependence is given by
\beq
g(z)=(1+1.11z) \sqrt{\Omega_M(1+z)^3 + \Omega_\Lambda} (1+z)^\delta \ .
\label{eq:gz}
\eeq
The extra mass dependence is given by
\beq 
f(M) = f(x,y) = 1+x+y- (1+x^{-1}+y^{-1})^{-1} \ ,
\eeq
where
\beq
x = \left(\frac{M}{10^{\MpkFid}} \right)^{- \alphaOneBest } \ , \quad
y = \left(\frac{M}{10^{\MpkFid}}\right)^{\alphaTwoBest} \ .
\eeq
In addition, $t$ is the cosmic time
\beq
t = 14.6 \int_z^{\infty} \frac{dz'}{(1+z')\sqrt{\Omega_M(1+z')^3 + \Omega_\Lambda}} \ {\rm Gyr}\ .
\eeq
Alternatively, one can use the fitting formula given in DM14, which is sufficiently accurate for $z>1$,
\beq
t = 17.5 (1+z)^{-1.5} \ {\rm Gyr}\ . 
\eeq

\section{Comparisons with other observations}\label{sec:comparisons}

We now compare our model predictions with other observations. We choose not to fit all observations simultaneously because of the different sources of systematic errors involved in them. In all the following calculations, we use 1\% of our MCMC chains to calculate the model predictions, and we plot the median as well as the 68\% and 95\% intervals for all quantities. In the main text, we only show the results of a single band or redshift bin for demonstration; the full comparisons can be found in Appendix~\ref{app:full}. This section focuses on direct observations from FIR/submillimetre surveys, including power spectra, number counts, and luminosity functions, while Section~\ref{sec:implications} focuses on derived quantities.

\subsection{Correlation between CFIRB and CMB lensing potential}\label{sec:lensing}
\begin{figure}
\centering
\includegraphics[width=\columnwidth]{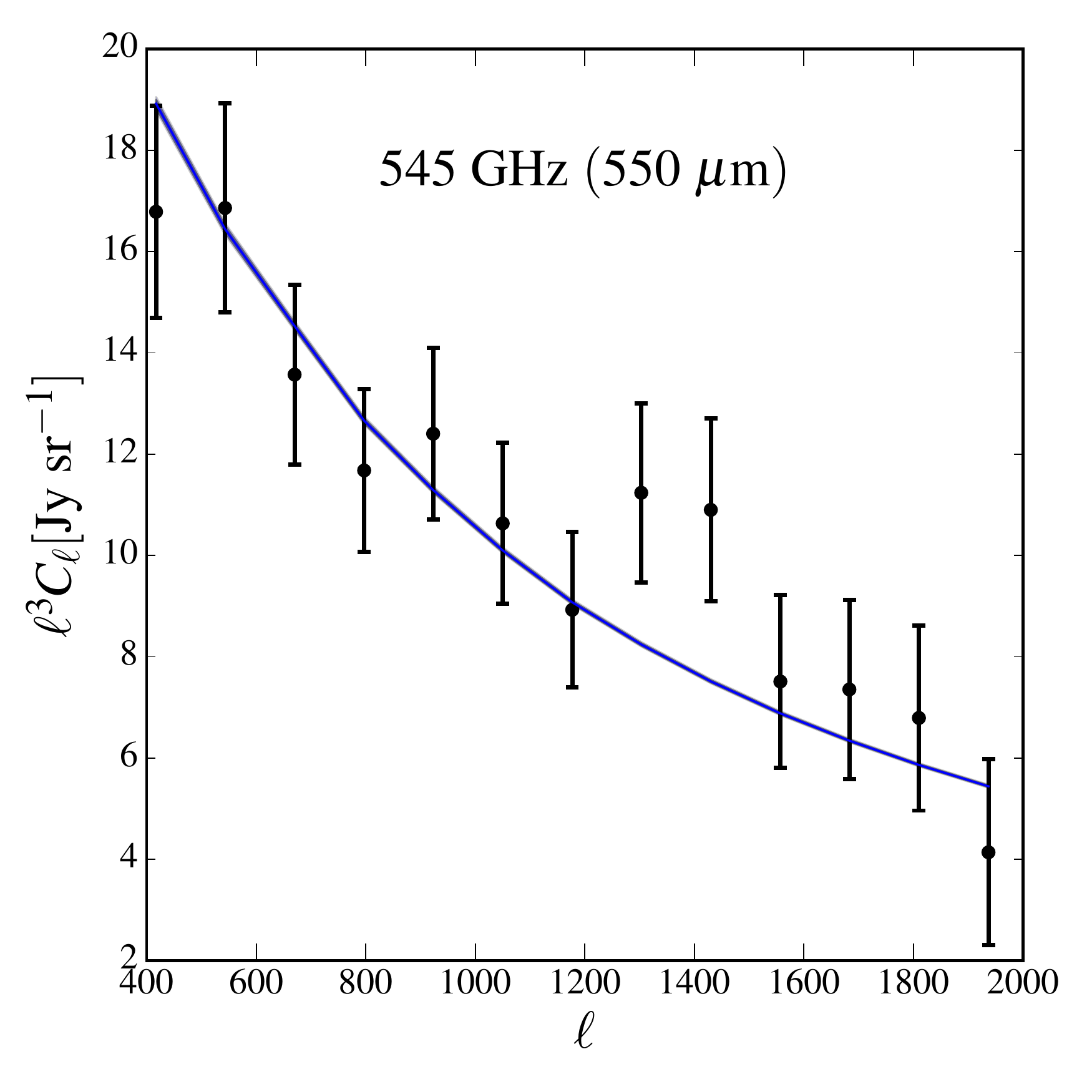}
\caption[Correlation between CFIRB and CMB lensing potential.]{Correlation between CFIRB and CMB lensing potential. The blue band is the prediction from our model, while the data points are from \protect\citet[see Fig.~\ref{fig:lensing_full} for all {\em Planck}-HFI bands]{Planck13XVIII}.}
\label{fig:lensing}
\end{figure}

\citet{Planck13XVIII} presented the first detection of the correlation between CFIRB and CMB lensing potential. The CMB lensing potential is dominated by haloes at $1\lesssim z \lesssim 3$ and is probed by the lower frequency bands of {\em Planck} (70 -- 217 GHz), while the CFIRB redshift distribution peaks at $1 \lesssim z \lesssim 2$ and is measured by the higher frequency bands of {\em Planck}. Therefore, the correlation between the two provides a powerful probe for the connection between dark matter and galaxies, as well as cross-check for systematics.

The cross power spectrum between the CMB lensing potential and CFIRB is given by
\beq
C_\ell^{\phi \nu } = \int_0^{\chi_*} B_\nu(z) \frac{3}{\ell^2} \Omega_M H_0^2 \left(\frac{\chi_* - \chi}{\chi_* \chi}\right) P_{\rm lin}\left(k=\frac{\ell}{\chi},z \right) d\chi \ , 
\eeq
where $\chi_*$ is the comoving distance to the last scattering surface, and $B_\nu(z)$ is given by equation (\ref{eq:Bz}) and is equivalent to $b_{\rm eff}(z) j_\nu(z)$.

Figs.~\ref{fig:lensing} and~\ref{fig:lensing_full} show that our model can recover the measurements presented in \citet{Planck13XVIII}. We note that the 68\% and 95\% intervals are very small because our model is constrained by the CFIRB spectra, which have much smaller error bars. Assuming that the IR luminosity is independent of halo mass, \citet{Planck13XVIII} applied a halo occupation distribution model and found that $\rm log_{10}(M_{\rm min}/\Msun) = 10.5\pm0.6$, where $M_{\rm min}$ is the minimum mass of a halo that hosts a central galaxy. \citet{Planck13XVIII} interpreted this mass scale as the characteristic mass for haloes hosting CFIRB sources; however, as we will see below in Section~\ref{sec:beff} and Fig.~\ref{fig:beff}, the effective galaxy bias consistent with this data set (as well as the CFIRB auto-correlation) corresponds to a halo mass of $10^{12.5} \ \Msun$ due to the mass dependence of SFR.

\subsection{Bolometric infrared luminosity Functions}\label{sec:LF}
\begin{figure}
\centering
\includegraphics[width=1\columnwidth]{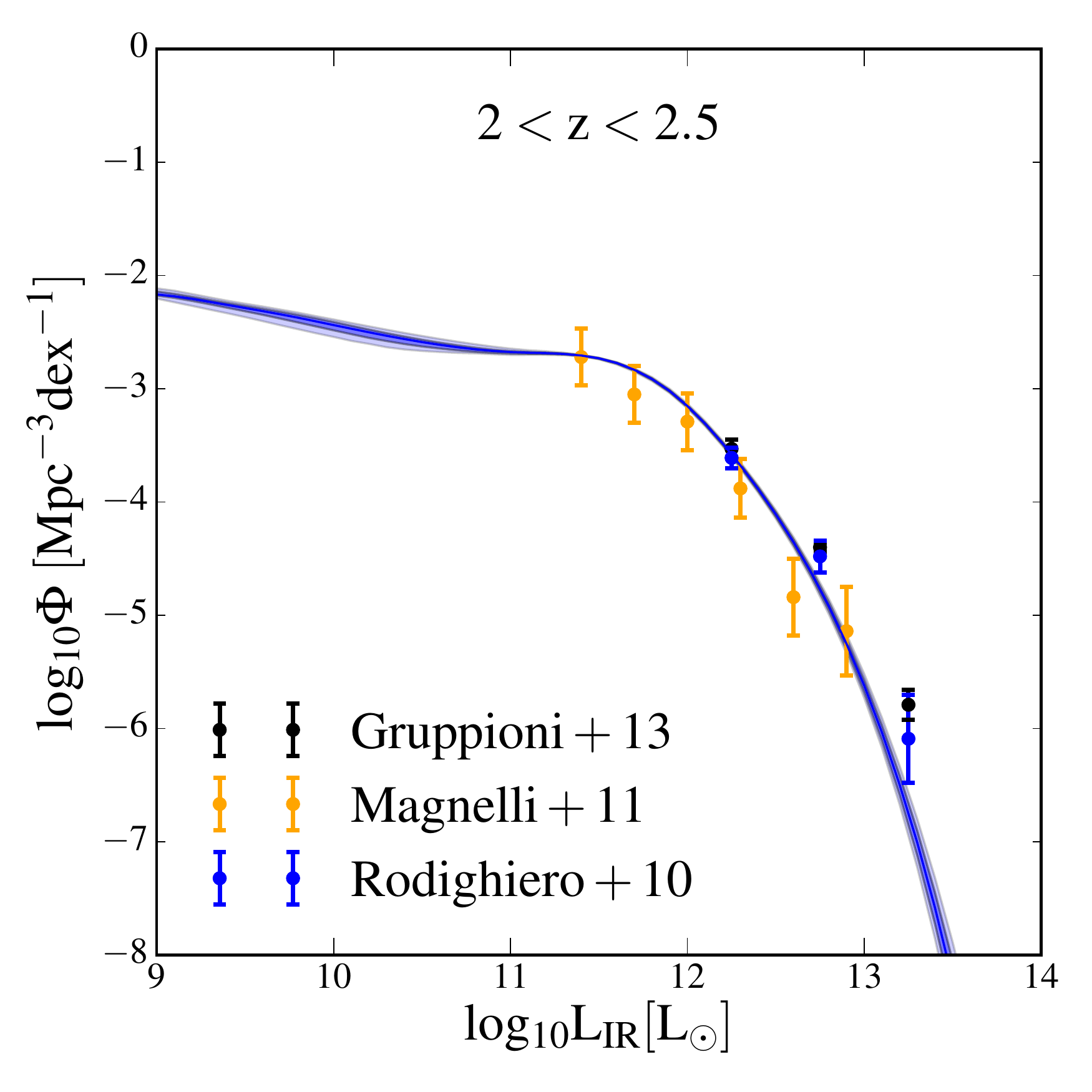}
\caption[IR luminosity functions.]{Bolometric infrared luminosity functions (8--1000\micron). The observational data sets include \protect\citet{Gruppioni13} from {\em Herschel}, as well as \protect\citet{Magnelli11} and \protect\citet{Rodighiero10} from {\em Spitzer} (see Fig.~\ref{fig:LF_full} for multiple redshift bins between $z=0$ and 4). }
\label{fig:LF}
\end{figure}

We assume that at a given $M$ and $z$, the natural logarithm of the IR luminosity ($\lnL$) of galaxy follows a normal distribution
similar to $\lnSnu$,
\beq
P(\lnL |M) = \frac{1}{\sqrt{2{\rm\pi}}\sigma}\exp\bigg[ -\frac{\big(\lnL - \avg{\lnL}\big)^2}{2\sigma^2} 
\bigg] \ .
\label{eq:pdf-L}
\eeq
Here $\sigma$ is the same as in equation (\ref{eq:pdf}). The luminosity function is given by
\beq
\frac{dn}{d\lnL}(z) = \int dM \frac{dn}{dM} P(\lnL|M) \ .
\eeq

We compare our model with the bolometric IR luminosity functions (integrating over 8--1000\micron) from the following publications:
\begin{itemize}
\item \citet[][table 6 therein]{Gruppioni13} based on {\em Herschel} (70--550 $\micron$), $0< z <4.2$. The galaxies are selected from PACS (70, 100, 160 \micron), and the SEDs are calibrated using SPIRE (250, 350, 550 $\micron$).
\item \citet[][table A6 therein]{Magnelli11} based on {\em Spitzer} (24 and 70 $\micron$), $1.3 < z < 2.3$.
They performed stacked analyses and derived the SED using the correlation between the luminosities at 24 and 70 \micron.
\item \citet[][table 5 therein]{Rodighiero10} based on {\em Spitzer} (8--24 $\micron$), $0 < z < 2.5$. The SED was derived using luminosities from optical to 24 $\micron$ and was thus not probing the peak of the dust emission. Nevertheless, their results are consistent with the results from \citet{Gruppioni13} based on longer wavelengths.
\item\citet[][table 2 therein]{LeFloch05} based on {\em Spitzer} 8 \micron, $0.3<z<1.2$. The bolometric luminosity was inferred from 24 $\micron$.
\end{itemize}

Fig.~\ref{fig:LF} shows the bolometric IR luminosity functions predicted from our model (see Fig.~\ref{fig:LF_full} for 11 redshift bins up to $z=4$). Since these data sets are based on slightly different redshift bins; we re-group these data points using the redshift bins in \citet{Gruppioni13} and compute the model at the middle of the bin. We note that all these observations are based on mid-infrared and use various SED templates to calculate the bolometric IR luminosity; therefore, they can suffer from different statistical and systematic uncertainties and do not necessarily agree with other. Therefore, we also expect that they will not necessarily agree with our model constrained by CFIRB. As can be seen in Fig.~\ref{fig:LF}, our model agrees with most of the data points but slightly under-predicts the bright end at high redshift. The scatter of the IR luminosity at a given mass ($\sigma$ in equation \ref{eq:pdf}), as constrained by CFIRB, determines the bright-end slopes of the IR luminosity functions.

\subsection{Number counts of FIR galaxies}\label{sec:NC}
\begin{figure}
\centerline{\includegraphics[width=1\columnwidth]{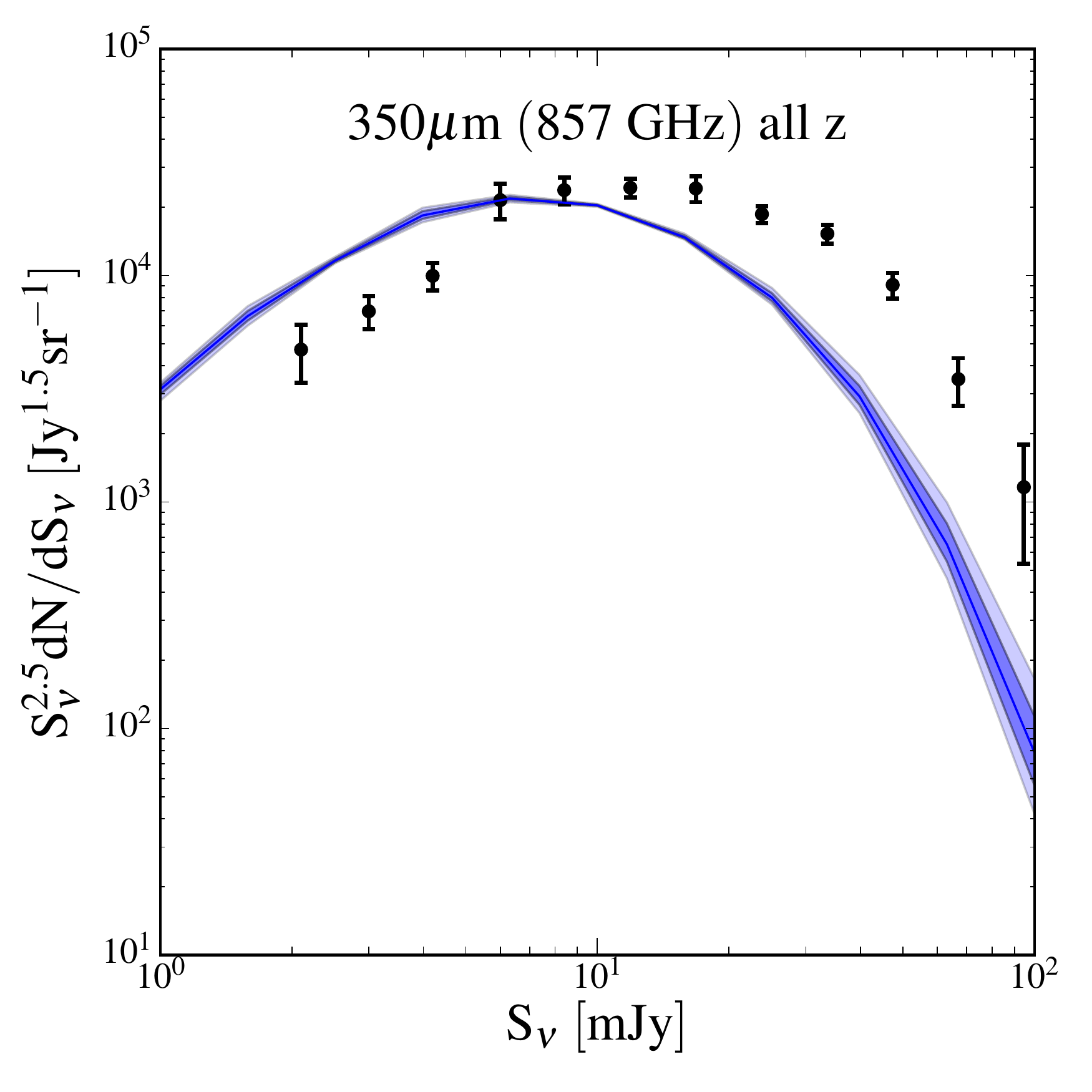}}
\caption[Number counts.]{Number counts of infrared galaxies. The data points are from \protect\citet{Bethermin12} based on {\em Herschel}-SPIRE. Our model under-predicts the bright source counts while over-predicts the faint source counts for all redshifts (see Fig.~\ref{fig:NC_full} for all {\em Herschel}-SPIRE bands and several redshift bins).}
\label{fig:NC}
\end{figure}
The number counts, also known as the flux density distribution function of infrared sources, is given by
\beq
\frac{dN}{dS_\nu}(z_1 < z < z_2 ) = \int_{z_1}^{z_2} dz \chi^2 \frac{d\chi}{dz} \frac{dn}{dS_\nu} \ .
\eeq
We compare our model with the deep number counts measured by \citet{Bethermin12} in the HerMES programme. These authors used the maps in 250, 350, and 500 $\micron$ in the COSMOS and GOODS-N fields observed by {\em Herschel}-SPIRE, and they used the catalogues of {\em Spitzer} 24 $\micron$ as priors for positions, flux densities, and redshifts. They provided the resolved number counts for $>$ 20 mJy and stacked number counts for between 2 and 20 mJy for several redshift bins.
 
Fig.~\ref{fig:NC} shows the comparison between our model (blue band) with the data points from \citet{Bethermin12}; the full comparison is presented in Fig.~\ref{fig:NC_full}. Our model under-predicts the number of bright sources and over-predicts the number of faint sources. We note that our model includes neither starburst galaxies nor strongly lensed galaxies, which can contribute to the bright end of the number counts functions. We also note that recently \cite{Bethermin17} show that the bright end of the number counts can be overestimated due to limited resolutions of the telescopes.

\subsection{Redshift distribution of CFIRB}\label{sec:dInu}
\begin{figure}
\centering
\includegraphics[width=1\columnwidth]{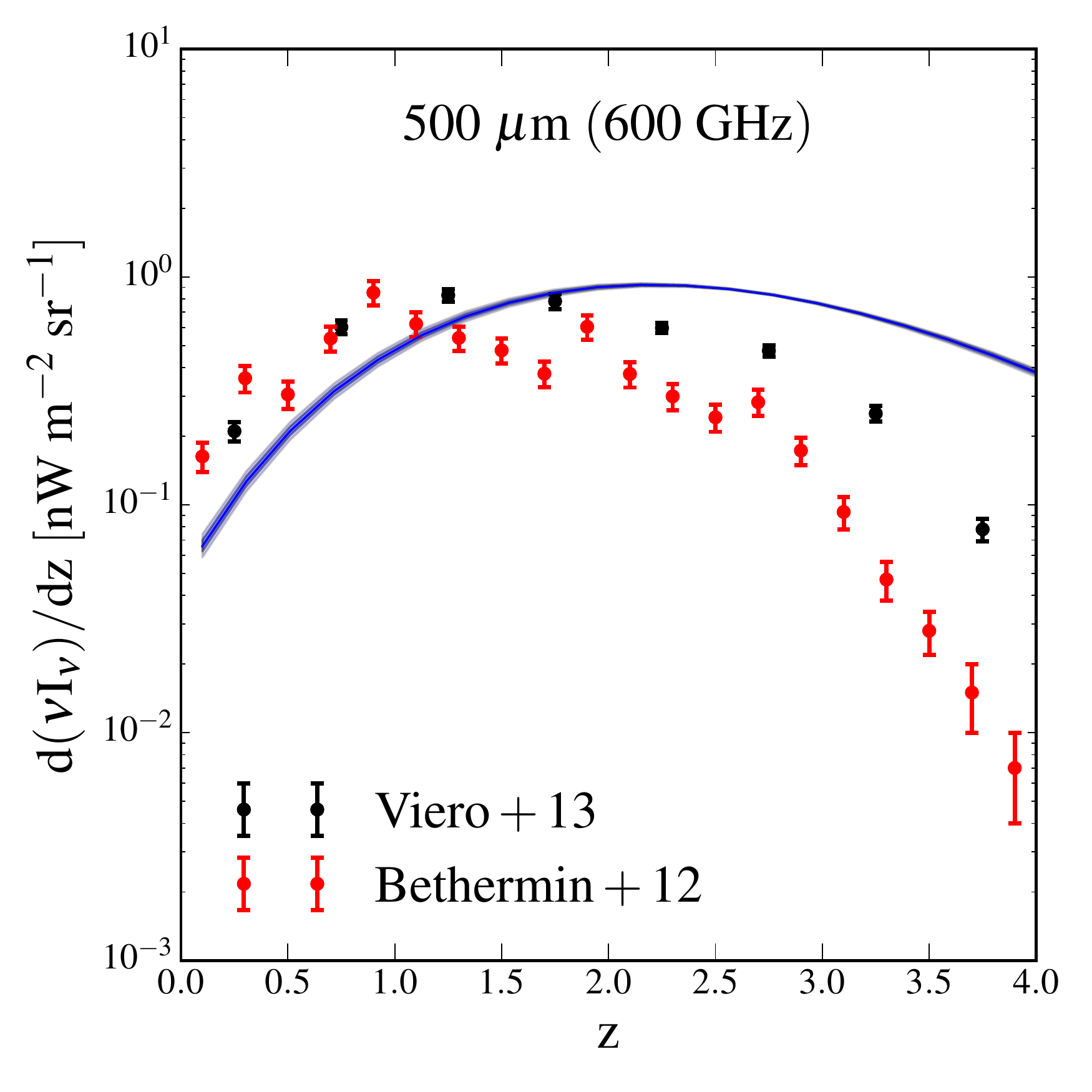}
\caption[Redshift distribution of CFIRB emission.]{Redshift distribution of CFIRB emission. The data points are from {\em resolved} sources in \protect\citet[black points, based on optical]{VieroMoncelsi13} and \protect\citet[red points, based on 24 $\micron$]{Bethermin12}, which serve as lower limits. Our model is above the data points for $z>1.5$ but is slightly lower for $z<1$ (see Fig.~\ref{fig:dInu_full} for all bands of {\em Herschel}--SPIRE).}
\label{fig:dInu}
\end{figure}
The redshift distribution of CFIRB emission is given by
\beq
\frac{dI_\nu}{dz} = \chi^2\frac{d\chi}{dz} \int dS_\nu \frac{dn}{dS_\nu} S_{\nu}
\eeq
We again compare our model with the data set from \citet{Bethermin12}, which was discussed in the previous section. 

Independently, \citet{VieroMoncelsi13} conducted a stacking analysis to quantify the fraction of CFIRB from galaxies resolved in optical. Specifically, they used the optical galaxy catalogue from the Ultra-Deep Survey fields from the UKIRT Infrared Deep Survey. Using the galaxy positions and photometric redshift, they performed stacking analyses on FIR maps, including the 250, 350, and 500 $\micron$ data from {\em Herschel}-SPIRE, and the 1100 $\micron$ data from AzTEC. With this analysis, they were able to separate the contribution of CFIRB from star-forming and quiescent galaxies in different stellar mass and redshift ranges. Their sample resolves 80\%, 69\%, 65\%, and 45\% of CFIRB in 250, 350, 500, and 1100 $\micron$, respectively. As mentioned in \citet{VieroMoncelsi13}, these measurements should be considered as lower limits, since optical catalogues can miss galaxies in FIR, either due to heavy dust obscuration or low intrinsic luminosity. The completeness also decreases rapidly with redshift. \citet{VieroMoncelsi13} also suggested that such measurements provide an effective way to break the degeneracies between redshift distribution, temperature, and halo bias.

Figs.~\ref{fig:dInu} and~\ref{fig:dInu_full} present the comparison between the redshift distribution of CFIRB from our model (blue band) and the results in \citet[red points]{Bethermin12} and \citet[black points]{VieroMoncelsi13}. Our model predicts higher differential intensity for $z>2$ than the data points, which should be considered as lower limits. If we use a lower differential intensity that is consistent with the data points, we will underestimate the total CFIRB intensity and clustering. On the other hand, our model predicts slightly lower differential intensity for $z<1$. 
This is consistent with what we saw in Fig.~\ref{fig:NC}, where our model also under-predicts the number counts for $z < 2$ observed by {\em Herschel}.

\subsection{CFIRB power spectrum from {\em Herschel}}\label{sec:CL_Herschel}
\begin{figure}
\centering
\includegraphics[width=\columnwidth]{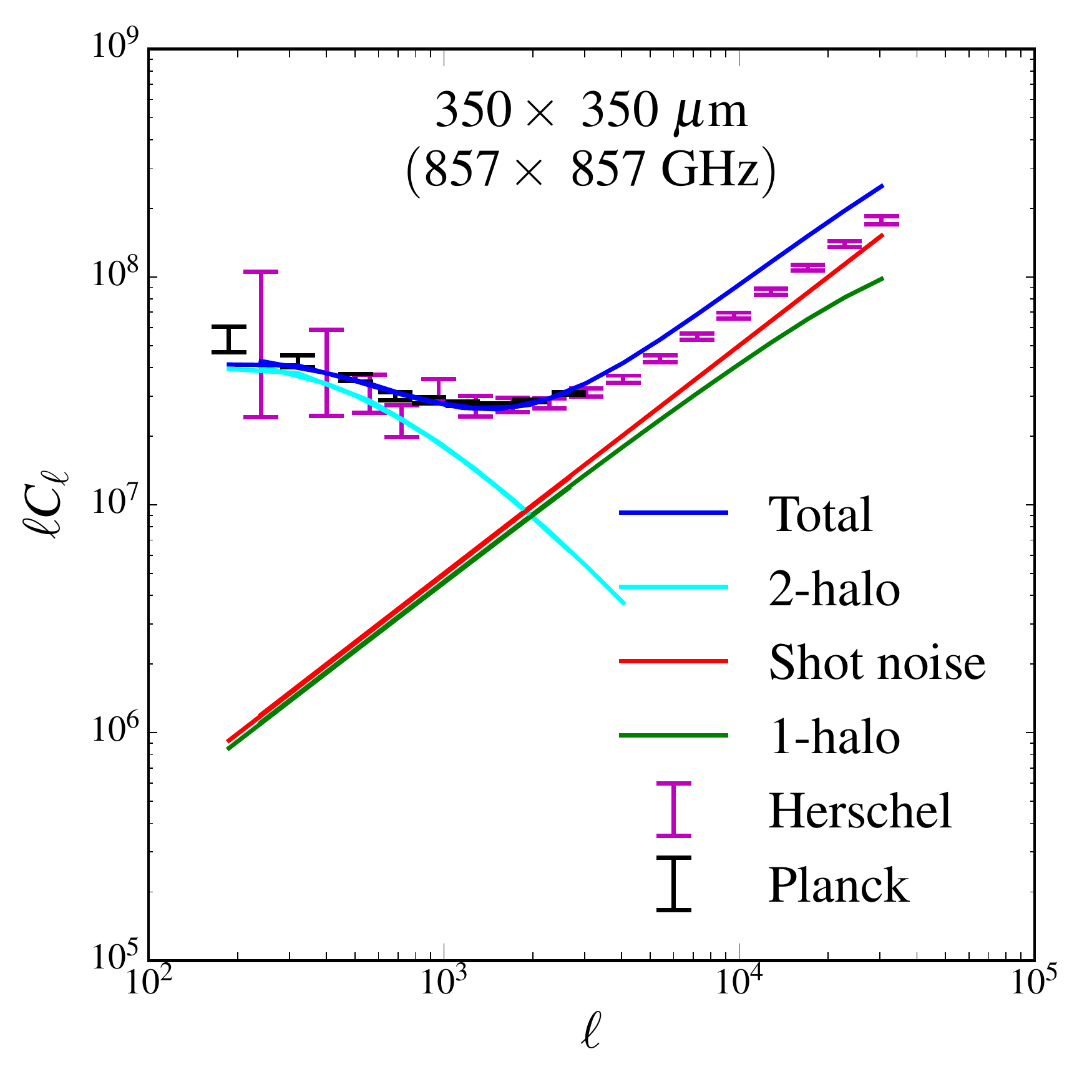}
\caption[Power spectra of {\em Herschel}.]{Comparison between our model and the power spectra from {\em Herschel} published in \protect\citet{Viero13}. Our model constrained by {\em Planck} overproduces the small-scale power when compared with {\em Herschel}. We show the power spectra of both {\em Planck} and {\em Herschel} in their common frequency 857 GHz (350 $\micron$) to illustrate the different angular scale and sizes of error bars (see Fig.~\ref{fig:CL_Herschel_full} for all frequencies for {\em Herschel}-SPIRE).}
\label{fig:CL_Herschel}
\end{figure}
Fig.~\ref{fig:CL_Herschel} shows the CFIRB power spectrum at 350 $\micron$ measured by {\em Herschel} \citep{Viero13}, compared with our model and the measurement of {\em Planck} in the same band. Fig.~\ref{fig:CL_Herschel_full} shows the comparison between our model and all frequencies of {\em Herschel}-SPIRE. The power spectra are based on the HerMES program, which covers 70 $\rm deg^2$ in 250, 350, and 500 $\micron$. The galactic cirrus was removed using the 100 $\micron$ maps from {\em IRAS}. Compared with the {\em Planck} data, the {\em Herschel} power spectra extend to smaller angular scales. As can be seen, our model over-predict the power for $\ell \gtrsim 4000$. The sum of the shot noise (red) and the 1-halo term (green) exceeds the data points. That is, the {\em Planck} power spectra favour higher clustering at small scales. Since the {\em Planck} power spectra has limited constraining power on small scale, extrapolating our results to small scales leads to this inconsistency with {\em Herschel} results.

\section{Implications of our model}\label{sec:implications}
Based on the constraints on parameters, we calculate various properties of dusty star-forming galaxies and compare them with observations. 

\subsection{IR luminosity--mass relation}\label{sec:LM}
\begin{figure}
\centering
\includegraphics[width=1\columnwidth]{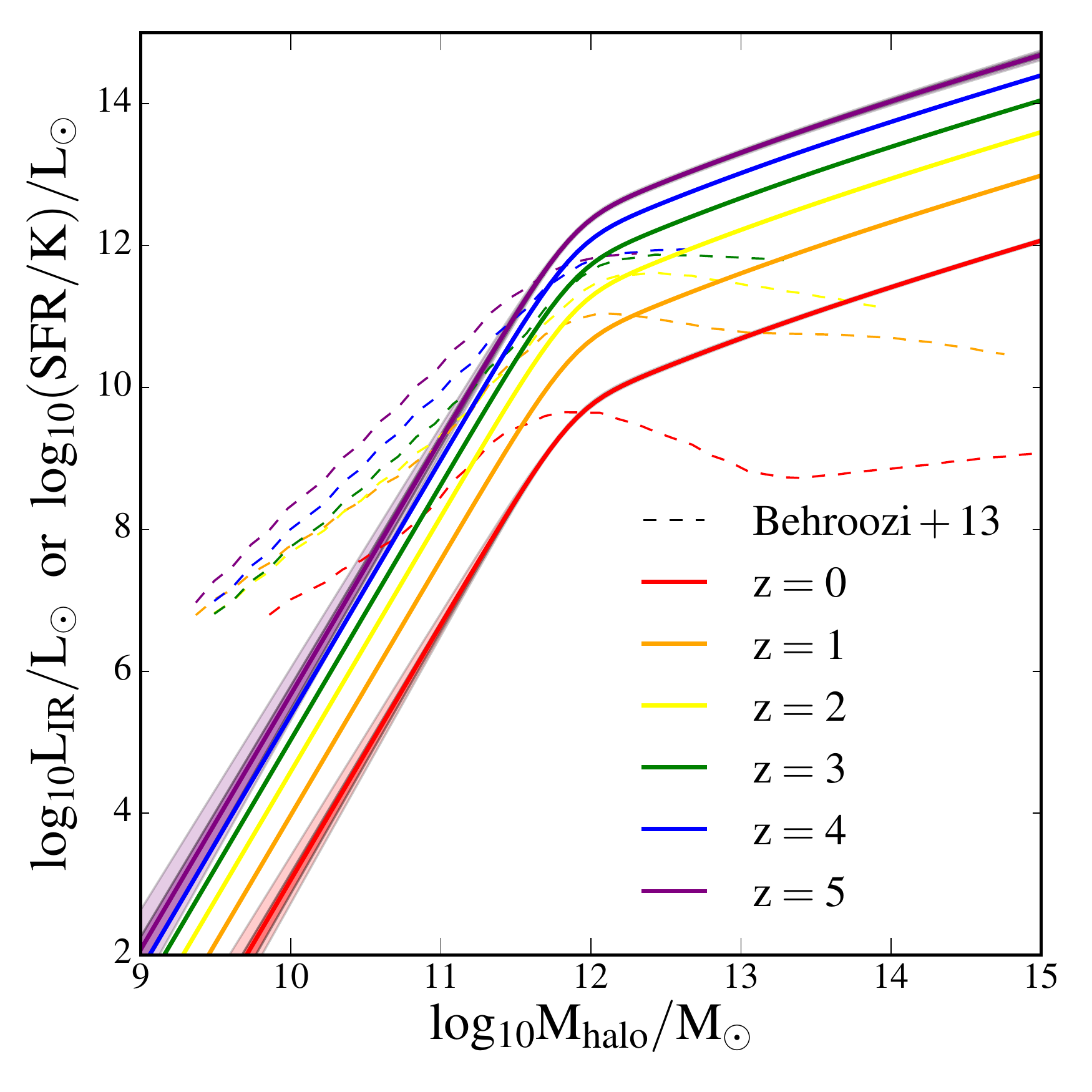}
\caption[IR luminosity--mass relation.]{Infrared luminosity versus halo mass in our model (solid curves); the function is provided in equation (\ref{eq:LM_final}). 
Comparing with the SFR constraints from UV/optical from \protect\citet[][dash curves]{Behroozi13} and assuming $\LIR = \SFR/K$, we find that the $\LIR$ at low-mass end is lower than expected from SFR, while the high-mass end requires higher $\LIR$ than expected from SFR.}
\label{fig:LM}
\end{figure}
Fig.~\ref{fig:LM} shows the mean relation between the infrared luminosity and the halo mass constrained by CFIRB (equation \ref{eq:LM_final}). The solid curves correspond to our model at various redshifts. The dash curves show the $\LIR$--$\Mh$ relation expected from the SFR from \citet{Behroozi13} and $\LIR = \SFR / K$.

For low-mass haloes, $\LIR$ is lower than the expectation from SFR. These low-mass haloes tend to be have low dust mass and thus lower IR luminosity given their SFR. For example, using hydrodynamic simulations with radiative transfer, \cite{Hayward14} have shown that low-mass galaxies are inefficient in absorbing UV photons, and inferring SFR from the IR luminosity can significantly underestimate the SFR for these galaxies \citep[also see, e.g.,][]{Jonsson06}. Using the data from HerMES, \cite{Heinis14} have found that galaxies with low stellar mass have lower dust attenuation, as well as lower IR excess (the ratio between $\LIR$ and $\LUV$); this confirms the findings in simulations that low-mass galaxies are inefficient in absorbing UV photons. Therefore, when converting SFR to $\LIR$, one should consider the mass dependence of dust attenuation. In \cite{Wu16b}, we show that a mass-dependent dust attenuation is crucial for recovering the observed CFIRB intensity and amplitude.

For massive haloes, the IR luminosity is significantly higher than what we expect from SFR. 
The CFIRB power spectra indicate a rather high galaxy bias that requires the contribution of FIR photons from massive haloes (see Section~\ref{sec:beff} below). If we use the $\LIR$--$\Mh$ relation from \citet{Behroozi13} in our halo model to calculate the power spectra, the amplitude of the CFIRB power spectra are too low regardless of the dust temperature used.

We note that, in addition to massive young stars, old stars can also heat the dust and contribute to FIR emission \citep[e.g.,][]{Groves12,Fumagalli14,Utomo14}. For example, using hydrodynamic simulations with radiative transfer, \citet{Narayanan15} have found that old stars can contribute to up to half of the IR luminosity. In addition, the heating from old stars contributes to a larger fraction of the IR luminosity for quiescent galaxies than for star-forming galaxies \cite[e.g.,][]{Fumagalli14}. Since these massive haloes tend to host quiescent galaxies, we expect that the contribution of heating from old stars is significant. 

On the other hand, dust-obscured AGN can also heat the dust and contribute the FIR emissions \citep[e.g.,][]{Alexander05,Lutz05, Yan05, LeFloch07,Sajina12}. However, the contribution from AGNs are expected to be low for massive galaxies; it has been shown that luminous AGNs are hosted by haloes of mass $10^{12}-10^{13}\ \Msun$
 \citep[e.g.,][]{AlexanderHickox12}. Therefore, AGNs are unlikely to be the main sources of the excess FIR emission.

The excess of FIR light for massive haloes has also be seen in previous publications. For example, \citet{Clements14} matched {\em Planck} sources and HerMES survey from {\em Herschel} and found four clumps consistent with galaxy clusters at $0.8<z<2.3$. They found that these cluster-like clumps have $\LIR = 3-70\times10^{12}\ \Lsun$; if one assumes that all the IR emissions are associated with star formation, such IR luminosities would imply an SFR of $600-10^4 \Msun yr^{-1}$. \citet{Narayanan15} used hydrodynamic simulations with radiative transfer to show that at $z\approx 2-3$, a dark matter halo of $10^{13}\Msun$ can have very high SFR ($500-1000\Msun yr^{-1}$). Such haloes can host groups of galaxies that are bright in submillimetre for a prolonged period due to constant gas infall. These findings suggest that there can indeed be IR-bright galaxies in massive haloes, which contribute the strong galaxy bias we find for CFIRB.

\subsection{Effective bias}\label{sec:beff}
\begin{figure}
\centering
\includegraphics[width=1\columnwidth]{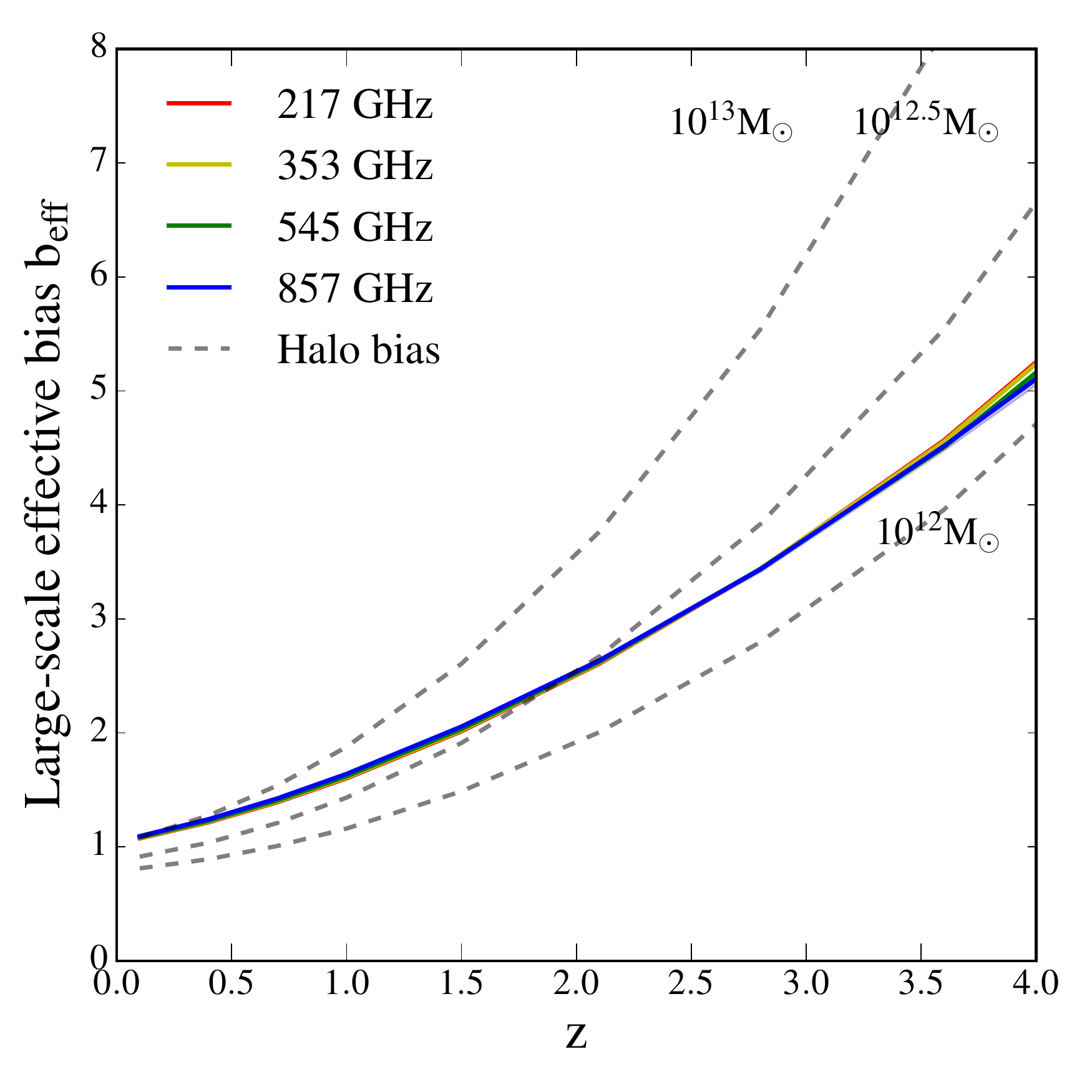}
\caption[Effective bias.]{The effective bias from our model, which is consistent with halo mass $10^{13} \Msun$ at $z=0$ and $10^{12.5}\Msun$ at $z=2$.}
\label{fig:beff}
\end{figure}
Fig.~\ref{fig:beff} shows the large-scale effective bias calculated from our model,
\beq
b_{\rm eff} = \frac{B_\nu(z)}{j_\nu(z)} \ ,
\eeq
where ${B_\nu(z)}$ and ${j_\nu(z)}$ are given by equations (\ref{eq:jz}) and (\ref{eq:Bz}). We note that since the SED depends on halo mass, the effective bias weakly depends on the frequency. For comparison, we show the bias of haloes of $\Mh = 10^{12,12.5, 13}\Msun$ as a function of redshift, using the fitting function from \citet{Tinker10}. As can be seen, our effective bias is consistent with haloes of mass $10^{13}\Msun$ at $z=0$ and $10^{12.5}\Msun$ at $z=2$. The CFIRB data favours a high galaxy bias and thus more contribution from haloes above $10^{12}\ \Msun$ .

An alternative explanation of this high galaxy bias could be that FIR galaxies represent biased environments, and the simple linear halo bias does not apply. It has been shown that the halo bias, in addition to its dependence on halo mass, can depend on formation time, concentration, and occupation \citep[e.g.][]{Wechsler06}. If FIR galaxies preferentially reside in haloes with recent major merger, or if the FIR luminosity and formation history are correlated, it might be possible to explain the high galaxy bias without invoking extra FIR sources in massive haloes. We will explore this in future work.

\subsection{Global SFR density}\label{sec:rho_SFR}

\begin{figure}
\centering
\includegraphics[width=1\columnwidth]{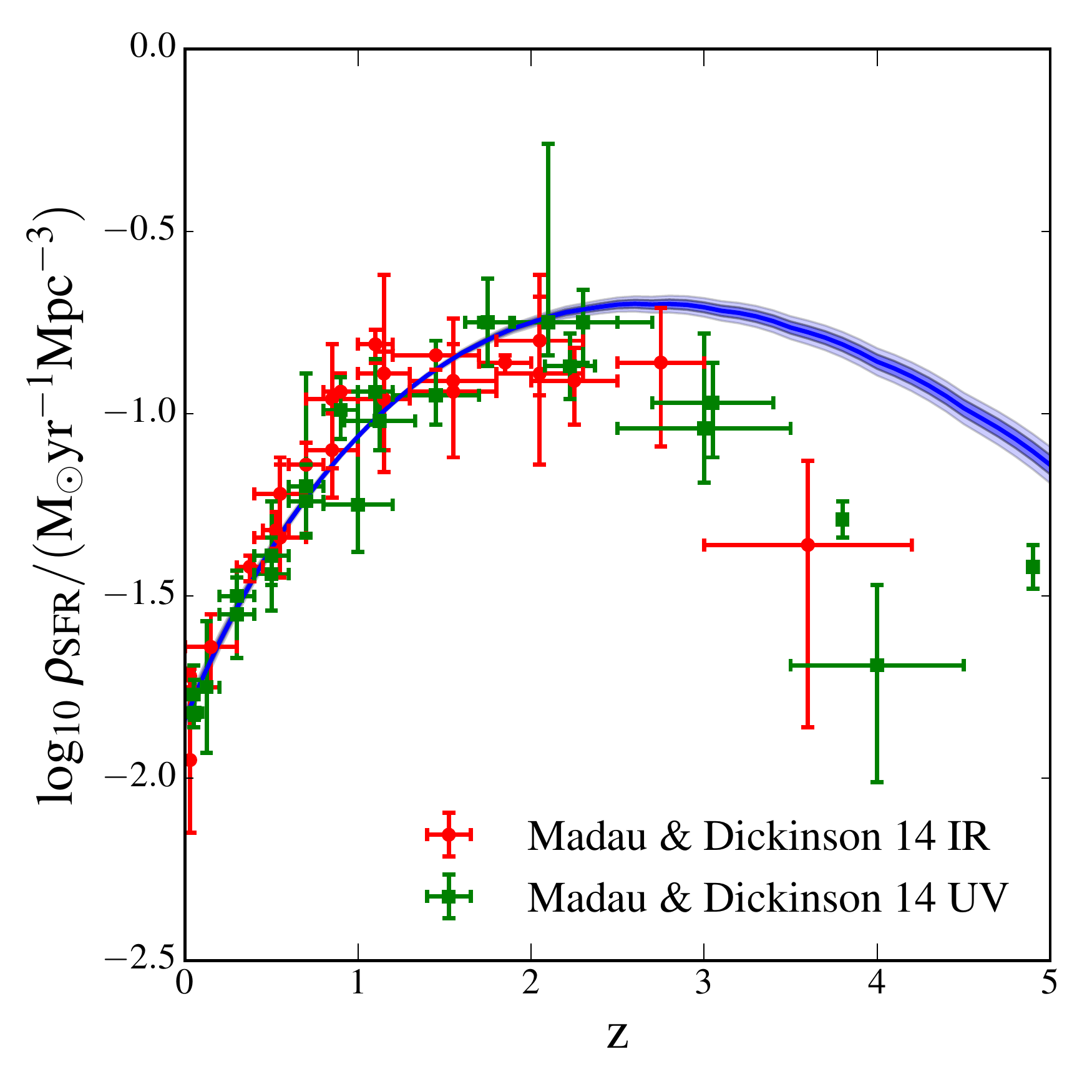}
\caption[Cosmic star formation history.]{Cosmic SFR density inferred from our model (blue band). Our model is consistent with the results from \protect\citet[][red and green points]{MadauDickinson14}.}
\label{fig:rho_SFR}
\end{figure}

Fig.~\ref{fig:rho_SFR} shows the SFR density based on our model,
\beq
\rho_{\SFR} (z) = K \int dM \frac{dn}{dM} \LIR(M,z) \ ,
\eeq
where $K$ is $1.7\times10^{-10}\Msun yr^{-1} \Lsun^{-1}$ \citep[][assuming Salpeter initial mass function]{Kennicutt98}.

We fit the four-parameter function proposed in \cite{MadauDickinson14} to our $\rho_{\SFR}$ \citep[also see][]{Robertson15}
\beq
\rho_{\SFR}(z)= a_p \frac{(1+z)^{b_p}}{1+[(1+z)/c_p]^{d_p}} {\rm M_\odot Mpc^{-3} yr^{-1}}
\eeq
where
\beqa
a_p &=0.0157_{-0.0004}^{+0.0003} \\
b_p &=2.51_{-0.03}^{+0.04} \\
c_p &=3.64_{-0.05}^{+0.04} \\
d_p &=5.46_{-0.09}^{+0.10} \ .
\eeqa
We note that these parameters are highly degenerate with each other.

For comparison, we plot the results based on UV and IR luminosity functions compiled by \citet[][table 1 and references therein]{MadauDickinson14}. The green points correspond to the results from FUV luminosity function (1500\AA) from {\em GALEX} and {\em HST} with corrections of dust attenuation. The red points correspond to the results from the IR luminosity function (8--1000\micron) from {\em IRAS}, {\em Spitzer}, and {\em Herschel}. We note that \citet{MadauDickinson14} re-computed the total luminosity density by extrapolating the best-fitting luminosity functions down to $0.03 L_*$ at each redshift from each publication. The faint-end slope and the dust extinction can therefore lead to significant uncertainties. They also cautioned that there is no robust measurements of SFR density for $z\gg 2$ due to the lack of robust selections. We also note that \citet{Robertson15} found results very similar to \citet{MadauDickinson14} when they added a few more UV results, extrapolated the observed UV and IR luminosity functions down to lower luminosities, and included the constraints of the integrated Thompson optical depth from \citet{Planck13cosmo}.

For $z<2$, our SFR agrees with the constraints from \cite{MadauDickinson14}. For high redshift ($z>3$), CFIRB does not provide strong constraints on the SFR, and the result is the extrapolation from low redshift; however, it is higher than UV constraints. We note that the halo model in \cite{Planck13XXX} also gave higher SFR density at high redshift, which could be related to their parametrization of redshift evolution \citep[also see][]{Serra16}. On the other hand, observations of gamma-ray bursts \citep[e.g.,][]{Kistler09} and UV background \citep[e.g.,][]{Mitchell-Wynne15} also hint at excess of SFR compared with the results from luminosity functions.

\subsection{Cosmic dust mass density}\label{sec:Omega_dust}

\begin{figure}
\centering
\includegraphics[width=1\columnwidth]{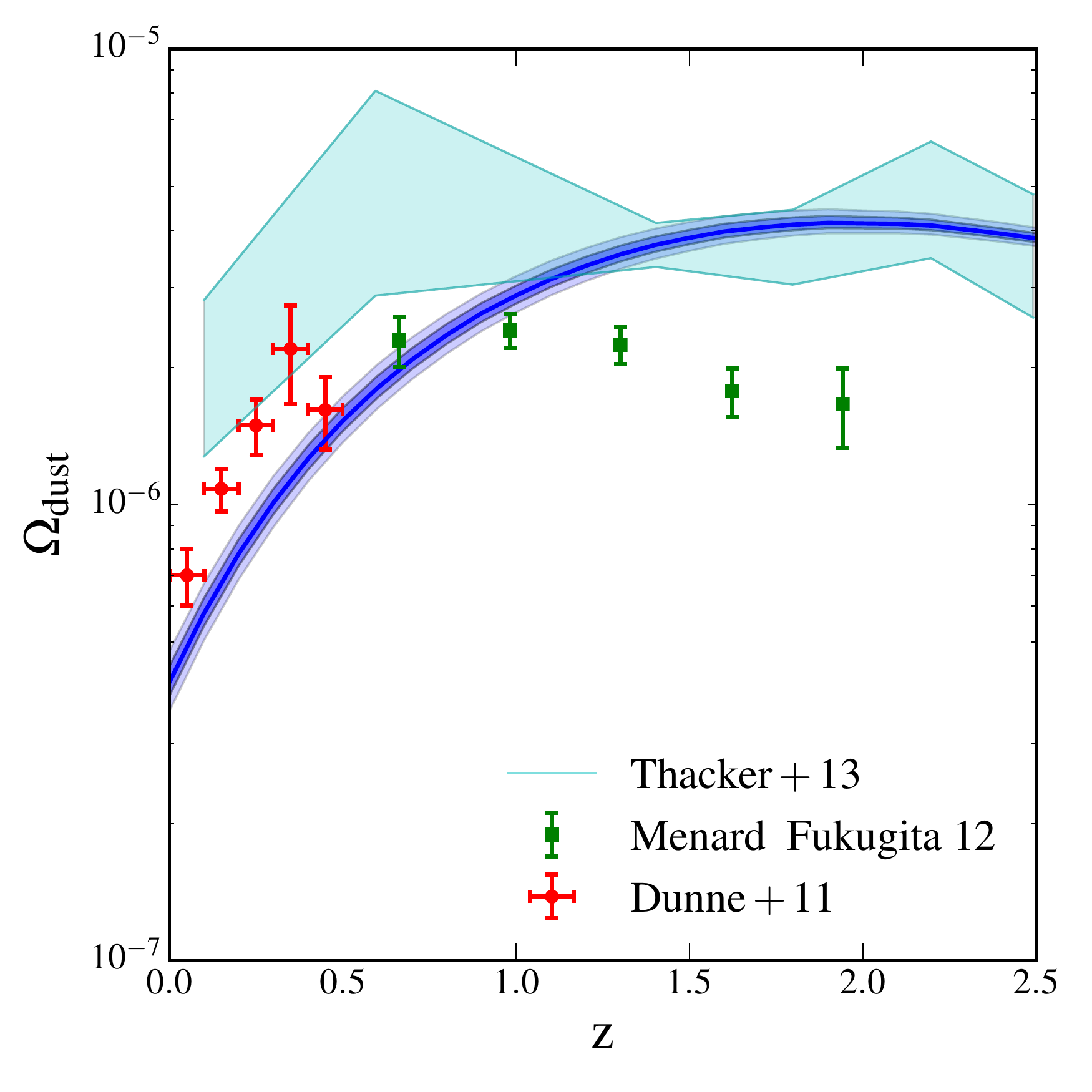}
\caption[Cosmic dust mass density.]{Cosmic dust mass density inferred from our model.
Compared with the results from \protect\citet{Thacker13} using CFIRB of H-ATLAS
and from \protect\citet{Dunne11} using the luminosity functions of H-ATLAS, our results are consistent at $z>1$ but are lower at $z<1$.
We note that the results from \protect\citet{Menard12} using Mg {\sc ii} absorbers serve as a lower limit of the dust in galactic haloes.}
\label{fig:Omega_dust}
\end{figure}

Fig.~\ref{fig:Omega_dust} shows the cosmic dust mass density calculated from our model.
The dust density is calculated by integrating over the halo mass function in physical units, 
\beq
\rho_{\rm dust} (z)= \int dz \frac{dn}{dM} M_{\rm dust}(M,z) \ .
\eeq
We express the dust mass density in unit of the critical density of the Universe,
\beq
\Omega_{\rm dust} (z)= \frac{\rho_{\rm dust} (z)}{\rho_{\rm crit} (z)} \ ,
\eeq
where 
\beq
\rho_{\rm crit}(z) = 2.775\times10^{11}
h^2 \bigg(\Omega_M (1+z)^3 +\Omega_{\Lambda}\bigg) \rm{ \Msun Mpc^{-3}} \ .
\eeq

For $z>1$, our results are consistent with the results of \citet{Thacker13} based on the CFIRB power spectra form H-ATLAS of {\em Herschel}. For $z<1$, our results are lower than \citet{Thacker13} and the low-redshift results of \citet{Dunne11}, which were derived from the luminosity functions of H-ATLAS. This is related to the fact that our model predicts lower number counts than those observed by {\em Herschel}. For comparison, we include the results using Mg {\sc ii} absorber from \citet{Menard12}. The dust mass density derived from Mg {\sc ii} serves as a lower limit for the dust associated with galactic haloes; the dust associated with galactic discs has been shown to be comparable to the dust associated with galactic haloes \citep{FukugitaPeebles04,Driver07}. Therefore, the total dust mass associated with galaxies is approximately twice of the values of the data points of \citet{Menard12}.

\subsection{Dust temperature and mass}

\begin{figure*}
\centering
\includegraphics[width=1\columnwidth]{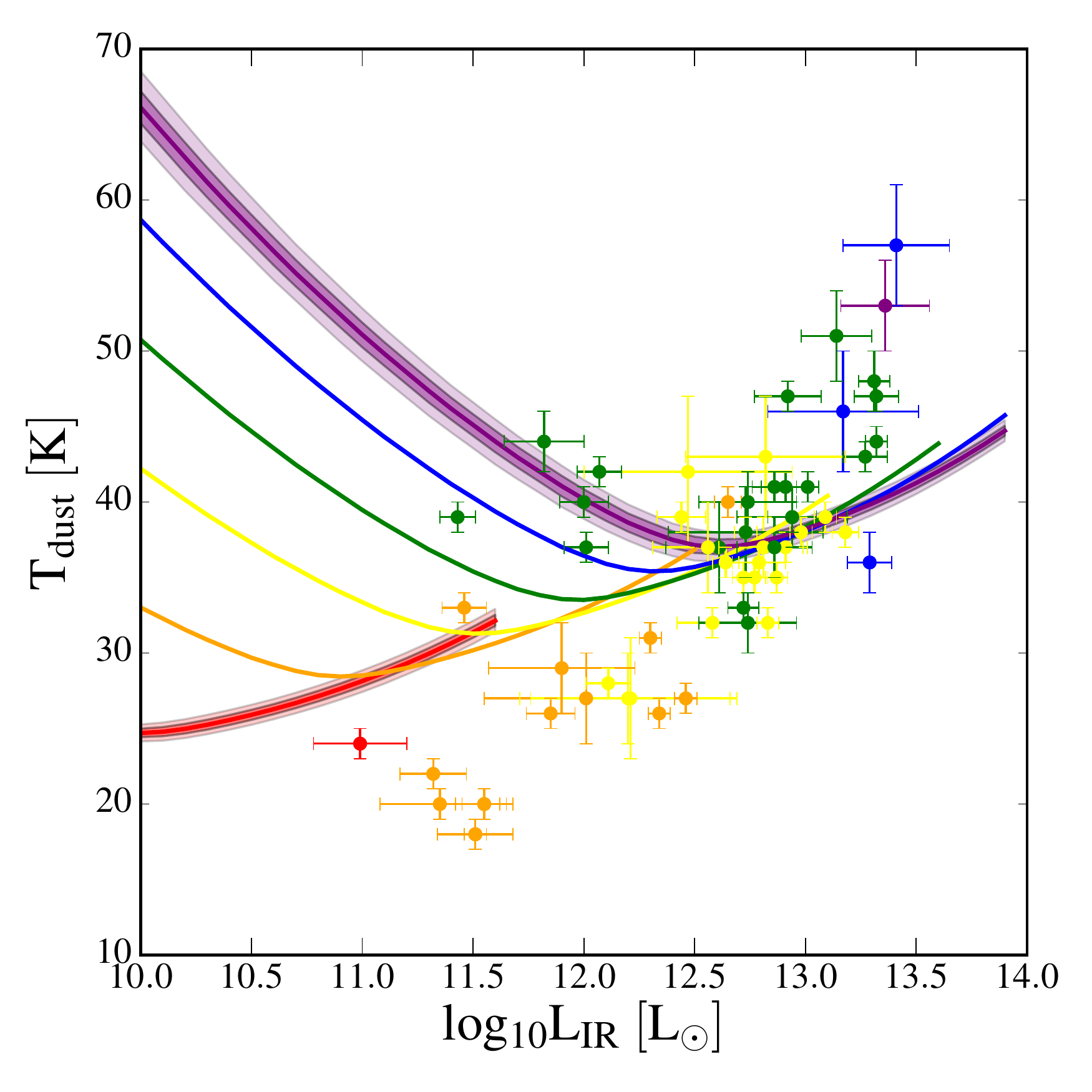}
\includegraphics[width=1\columnwidth]{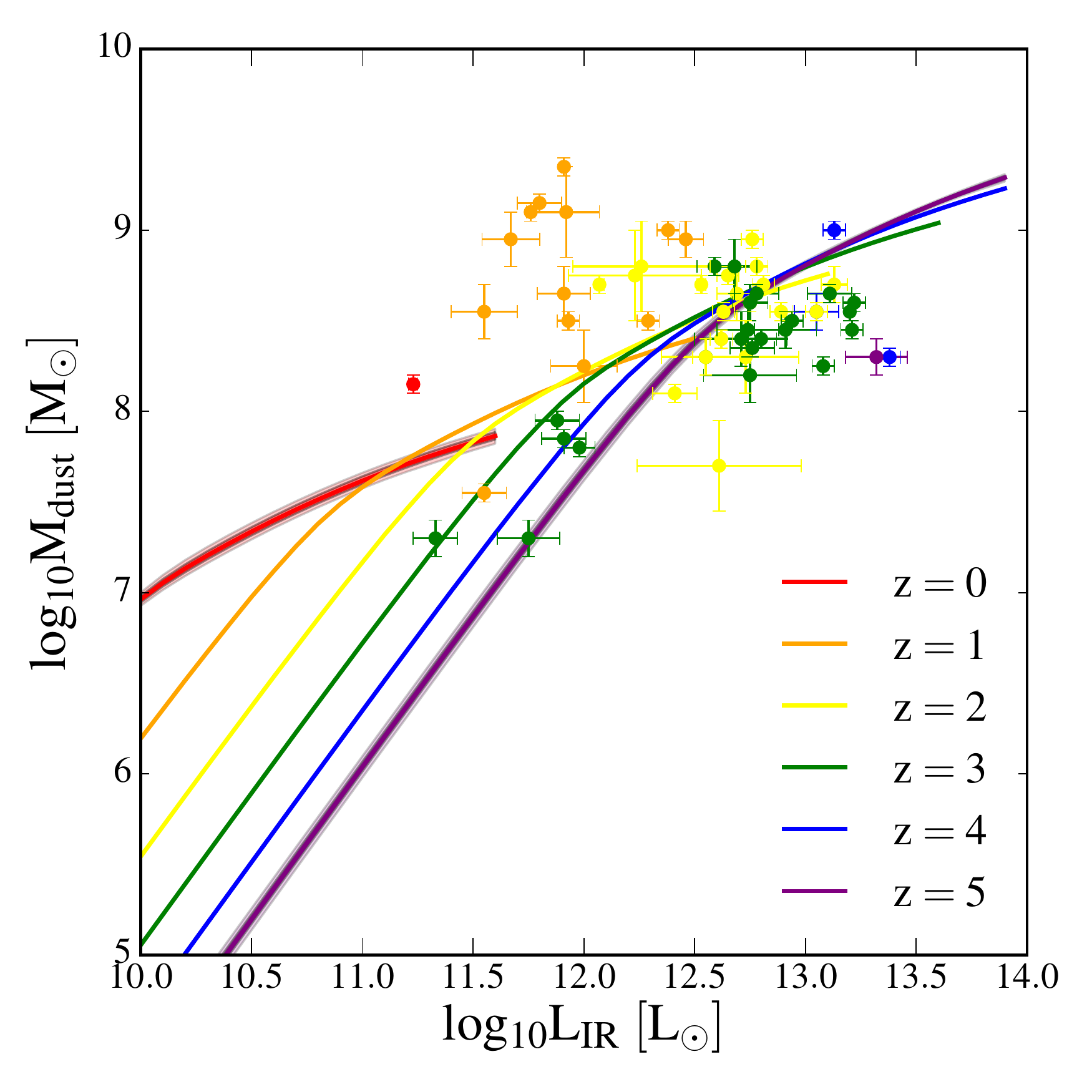}
\caption[Dust temperature and mass.]{Dust properties from our model 
compared with observations of 61 SMGs observed by {\em Herschel} \protect\citep{Magnelli12}. We only show the the model uncertainties at $z$ = 0 and 5, and the uncertainties at other redshifts are very similar. We note that this figure is mainly for demonstrating the orders of magnitude because of the complex selection function involved. {\bf Left}: Our model predicts higher dust temperature than the observation. {\bf Right}: Our mean dust mass is slightly lower than the observation.}
\label{fig:Mdust}
\end{figure*}

Fig.~\ref{fig:Mdust} shows the dust properties from our model. The left-/right- hand panel corresponds to dust temperature/mass versus IR luminosity at various redshifts, shown by different colours. Our model predicts non-monotonic relations with $\LIR$; $\Md$ tends to be low at both the bright and faint ends, while $\Td$ tends to be high at both ends. This can be understood through the mass dependence of the mass-loading factor. In our model, the dust mass is given by
\beq
\Md \propto \frac{1}{(1-R+\eta)^2}
\eeq
(see equations \ref{eq:Mdust} and \ref{eq:Mdust_final}). The high mass-loading factor for both high- and low-mass haloes leads to strong mass outflow and thus low dust mass. In addition, under the assumption of local thermal equilibrium, the dust temperature depends on the ratio between $\LIR$ and $\Md$,
\beq 
\Td \propto \bigg(\frac{\LIR}{\Md}\bigg)^{1/(4+\beta)} 
\propto ( {1-R+\eta} )^{1/(4+\beta)}
\eeq
(see equations \ref{eq:Tdust} and \ref{eq:Tdust_final}). Therefore, at a given redshift, haloes at both high- and low-mass ends tend to have high dust temperature due to the high mass-loading factor. 

We compare our results with the observational results in \citet[][M12 thereafter]{Magnelli12}, which include 61 submillimetre galaxies (SMGs) selected from ground-based observations and observed with PACS and SPIRE instruments onboard {\em Herschel}. We caution that this comparison is mainly for demonstrating the range of values, as the observations of SMGs tend to select merger-driven starbursts and has incomplete coverage for the main-sequence galaxies. As stated in M12, for high IR luminosity, the sample is representative of the entire SMG population, but these galaxies tend to be associated with merger-driven starbursts; on the other hand, for low IR luminosity, the sample tends to bias towards low redshift and colder dust. M12 concluded that approximately half of the sample is consistent with the merger-drive starbursts, while the other half is consistent with the main sequence of stellar mass and SFR. That is, this sample may not be relevant for the galaxies contributing to the CFIRB. 

The left-hand panel of Fig.~\ref{fig:Mdust} shows the relation between $\Td$ and $\LIR$ predicted from our model. The correlation between $\LIR$ and $\Td$ has been known for SMGs \citep{Chapman05,Hwang10,Hayward12}. In M12, $\Td$ and $\LIR$ are derived from fitting the SED to a modified blackbody with a single dust temperature, with $\beta = 1.5$. We note that M12 used 40--120 $\micron$ to calculate $L_{\rm FIR}$ and assumed $\LIR = 1.91 L_{\rm FIR}$. The dust temperature from M12 is lower than ours for $z<2$. This may reflect the SMG selection tends to bias towards low-redshift, low-temperature galaxies. In our model, the trend is reversed for faint galaxies; since we require strong feedback to suppress the SFR for low-mass haloes, this feedback also suppresses the dust mass and increases the dust temperature. 

The right-hand panel of Fig.~\ref{fig:Mdust} shows the relation between $\Md$ and $\LIR$ from our model, as well as the measurements in M12. To derive the dust mass, M12 assumed a power-law distribution of dust temperature and fit the SED. Our model is consistent with M12. Nevertheless, M12 showed higher dust mass for $z<2$, and this difference is related to the lower dust temperature seen in the left-hand panel.

\section{Discussion}\label{sec:discussion}

In this work, we show that the gas regulator model provides a qualitative description for the CFIRB power spectra but is unable to produce all the details in observations. In this section, we discuss the limitations of our implementation of the gas regulator model and possible improvements.

In our implementation, we assume that most of the parameters are time-independent and mass-independent, and we incorporate the mass-dependence in the mass-loading factor (equation \ref{eq:etaM}) and the extra time-dependence in SFR (equation \ref{eq:fz}). These parametrizations attempt to capture the effects of feedback, but they do not capture the detailed physics and thus cannot reproduce observations perfectly. The effective mass-loading factor $\eta$, the accretion of gas $\fga$ and stars ($1-\fga$), and the return fraction $R$ can all have non-trivial time and mass dependence. Capturing the time and mass dependence accurately would require hydrodynamic simulations or semi-analytic models. The limitations of the gas regulator model have also been demonstrated in the literature. For example, DM14 have shown that their fiducial model systematically under-estimates the specific SFR at $1 < z < 4$. In our work, we use a few free parameters but are still unable to fit the data perfectly. 

\cite{KrumholzDekel12} implemented a metal-dependent SFR to take into account the fact that at high redshift, low-mass galaxies tend to have low metallicities and are unable to sustain a cool gas reservoir. Therefore, the SFR for low-metallicity galaxies is suppressed and is lower than what we would expect from the gas accretion rate. In our model, this effect is mimicked by the high effective mass-loading factor for low-mass galaxies. Our model qualitatively captures such trend; however, in principle, the SFR should be modelled self-consistently given the metallicity and dust. 

Furthermore, in our model we assume that all galaxies follow a simple modified body SED with the dust temperature calculated by assuming thermal equilibrium. This assumption is too simplistic and may be the reason why we have significantly worse fit in the 217 GHz band.

Our model does not include starburst galaxies, which contribute to $\sim$10 per cent of the cosmic SFR density at $z\sim 2$ \citep{Rodighiero11,Sargent12} and are expected to have negligible contribution to CFIRB \citep{Shang12,Bethermin13}.  Including the starburst galaxies could increase the bright end of the luminosity functions and number counts. However, an extra component for starburst would boost the power spectrum in the same way as a higher gas accretion rate would, and breaking such degeneracy would require a joint fit to the bright end of the luminosity functions.

\section{Summary}\label{sec:summary}

We apply the gas regulator model of galaxy evolution to describe dusty star-forming galaxies across cosmic time. We fit the model to the CFIRB power spectra observed by {\em Planck}. We compare our model predictions with the total CFIRB intensity measured by {\em COBE}, the correlation between CFIRB and CMB lensing potential measured by {\em Planck}, the bolometric IR luminosity functions up to $z=4$ from {\em Herschel} and {\em Spitzer}, and the total number counts from {\em Herschel}. The implications of our model are summarized as follows:

\begin{itemize}

\item The CFIRB power spectra favour a strong clustering of FIR galaxies. At $z=0$ ($z=2$), the large-scale galaxy bias is equivalent to the bias of dark matter haloes of mass $10^{13}$ ($10^{12.5}$) $\Msun$. This galaxy bias is consistent with the correlation between CFIRB and CMB lensing potential.

\item The luminosity--mass relation from our model indicates that for massive haloes, the IR luminosity is higher than expected from the SFR constrained by UV and optical. This result is consistent with the high galaxy bias we have found. This excess in IR luminosity for massive haloes may come from the dust heated by old stellar populations.

\item In our model, the luminosity--mass relation for low-mass haloes is lower than expected from the SFR constrained by UV and optical. These low-mass galaxies tend to be inefficient in absorbing UV photons, and their FIR emissions can underestimate the true SFR.

\item Our model under-predicts the bright source counts of {\em Herschel}, slightly under-predicts the differential CFIRB intensity of {\em Herschel} for $z<1$, and over-predicts the CFIRB power spectra of {\em Herschel} at small scales.

\item The cosmic star formation history from our model agrees with the recent compilation of \citet{MadauDickinson14} at $z<2$ but shows an excess at higher redshift. In addition, the total dust mass density across cosmic time is consistent with the results from {\em Herschel} CFIRB at $z>1$, while it is lower than the results from IR luminosity functions at $z<1$. 

\item Compared with SMGs selected from ground-based surveys, the galaxies in our model tend to have higher dust temperature ($T_{\rm dust} \gtrsim 25$ at $z=0$ and increases with redshift) and lower dust mass.

\end{itemize}

Our theoretical framework provides a simple, physically-motivated way to compare different FIR observations. It can be generalized to compute the foreground for various intensity mapping experiments. Our framework will also be useful for optimizing the survey designs and strategies for future FIR surveys. For example, the next generation CMB experiments, such as PIXIE \citep{Kogut11} and CORE \citep{CORE15}, will provide larger frequency coverage and/or higher angular resolution and sensitivity than {\em Planck} and will be able to provide better measurements for the CFIRB anisotropies as well as individual sources. In \cite{Wu16c}, we investigate the constraining power from future CFIRB experiments. The Far-IR Surveyor, which is currently explored by NASA\footnote{http://asd.gsfc.nasa.gov/firs/}, will reveal many more properties of dusty star-forming galaxies.

\section*{Acknowledgements}
We thank Chris Hayward, Lorenzo Moncelsi, Jason Sun, and Marco Viero for helpful discussions. H.W.\ acknowledges the support by the U.S.\ National Science Foundation (NSF) grant AST1313037. The calculations in this work were performed on the Caltech computer cluster Zwicky, which is supported by NSF MRI-R2 award number PHY-096029, and on the Piz Dora cluster of the Swiss National Supercomputing Centre. Part of the research described in this paper was carried out at the Jet Propulsion Laboratory, California Institute of Technology, under a contract with the National Aeronautics and Space Administration.

\bibliographystyle{mnras}
\bibliography{/Users/hao-yiwu/Dropbox/master_refs}

\appendix
\section{Summary of parameters}

\begin{figure*}
\centering
\includegraphics[width=2\columnwidth]{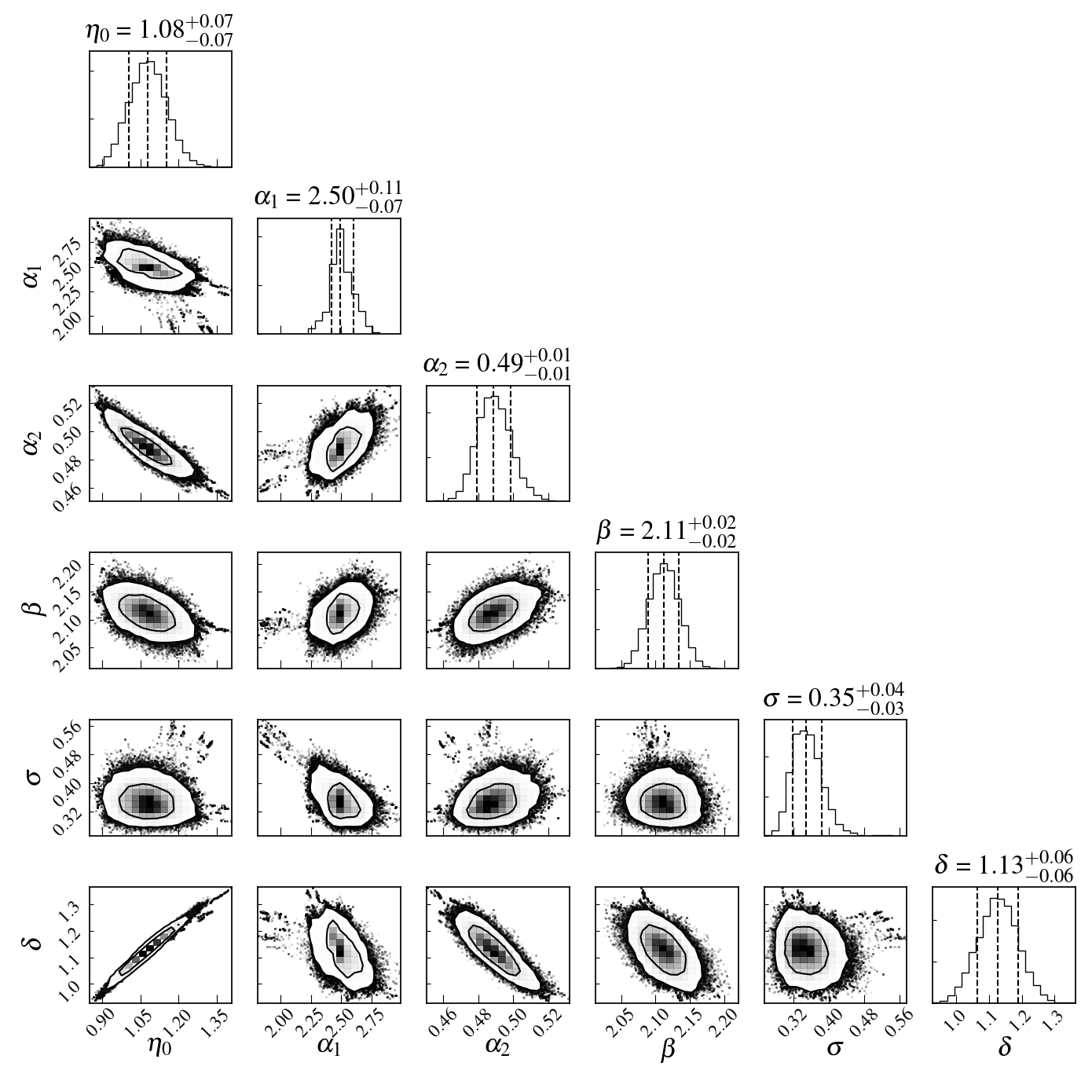}
\caption[MCMC]{The 68\% and 95\% constraints of our model parameters. The diagonal panels show the posterior distribution and the 68\% constraint of each parameter.}
\label{fig:MCMC}
\end{figure*}

Table~\ref{tab:cor} shows the correlation matrix of these parameters. Fig.~\ref{fig:MCMC} shows the 1-D and 2-D posterior distributions from the MCMC chains, which use the {\sc corner} software \citep{corner}.

\begin{figure*}
\centering
\includegraphics[width=2\columnwidth]{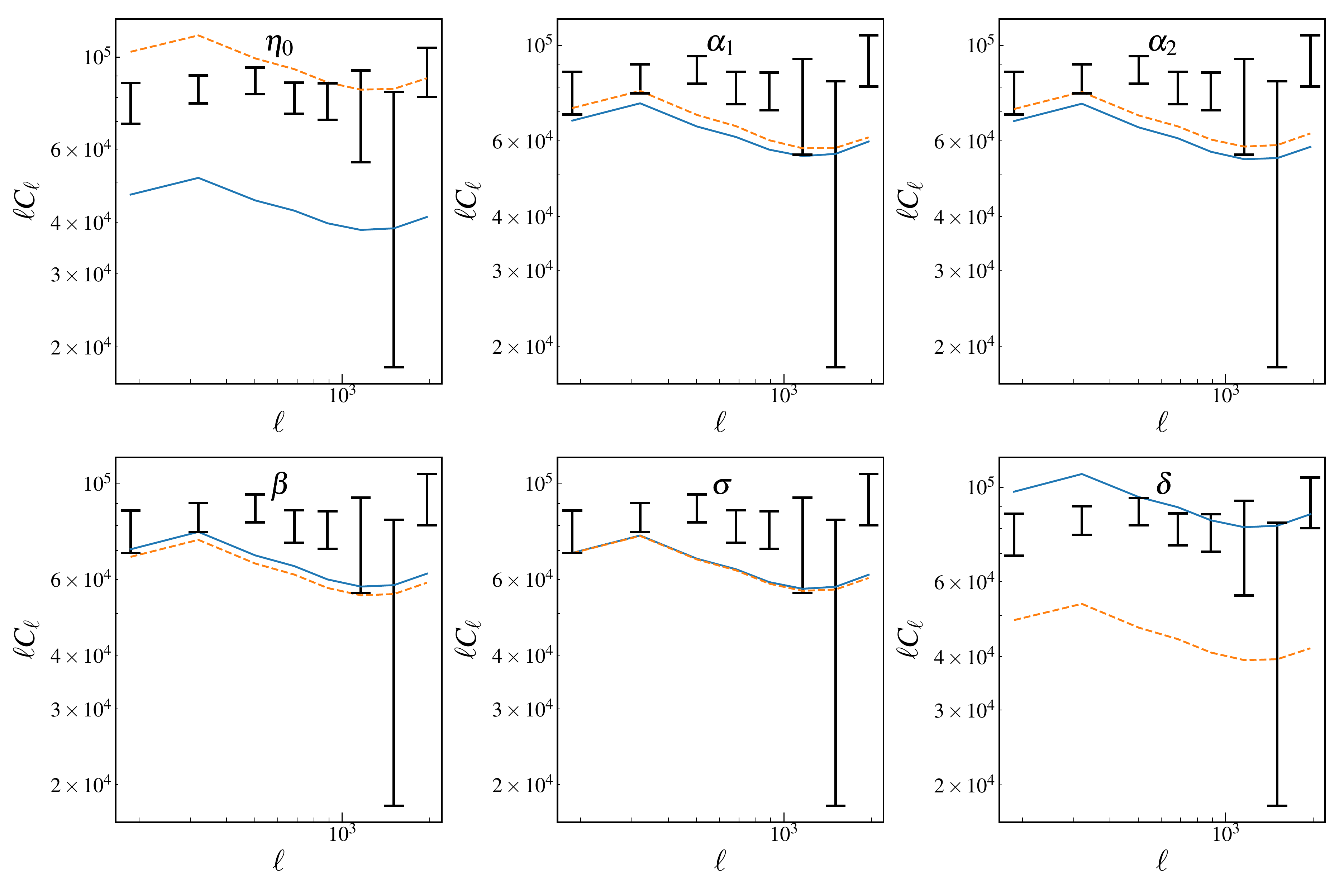}
\vspace{-0.7cm}
\caption[]{Sensitivity of angular power spectra (217 GHz) to model parameters. In each panel, the solid and dash curves correspond to increasing and decreasing the model parameters by $2\sigma$.}
\label{fig:CL_217_sensitivity}
\end{figure*}

\begin{figure*}
\centering
\vspace{-0.5cm}
\includegraphics[width=2\columnwidth]{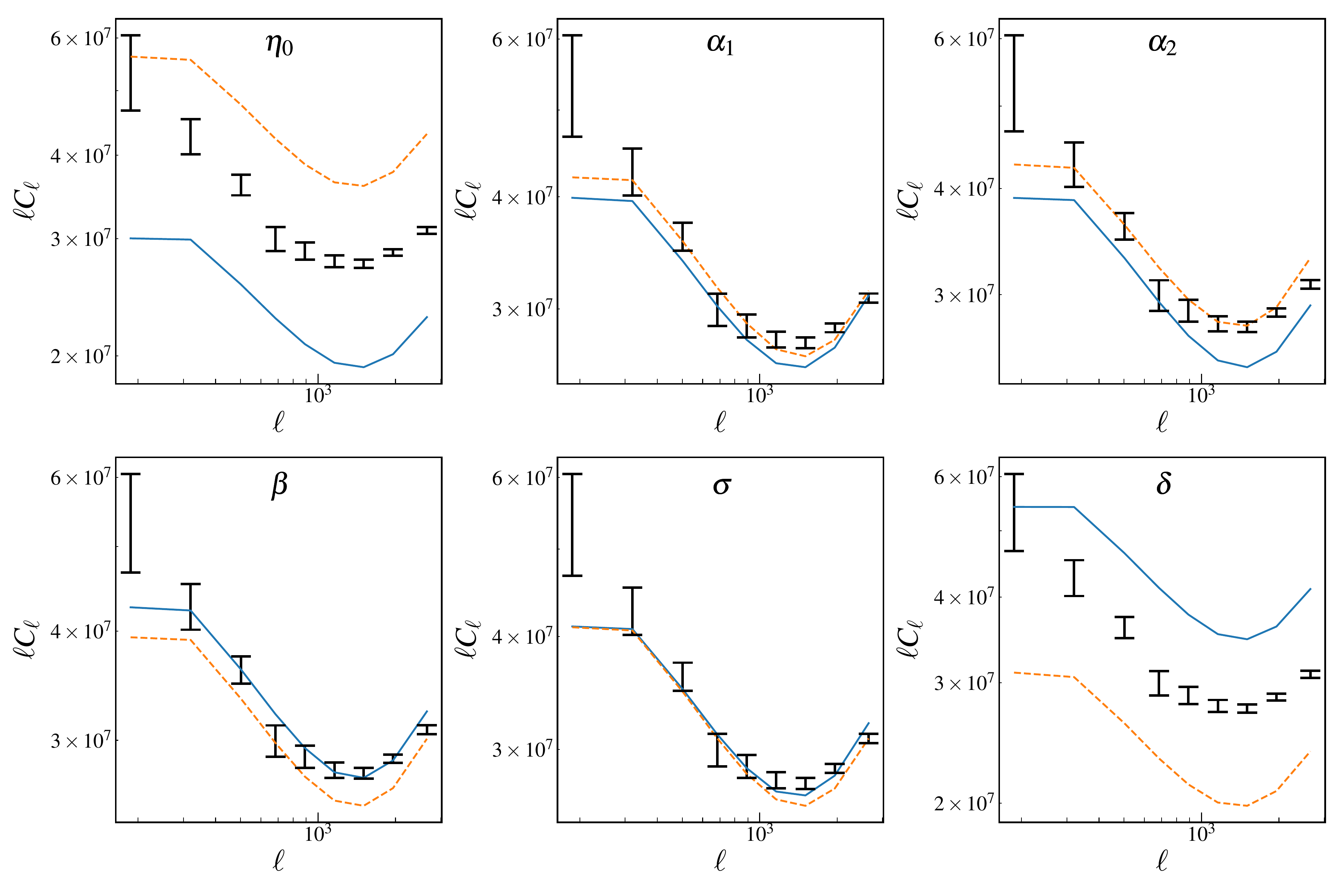}
\caption[]{Same as Fig.~\ref{fig:CL_217_sensitivity} but for 857 GHz.}
\label{fig:CL_857_sensitivity}
\end{figure*}

Figs.~\ref{fig:CL_217_sensitivity} and \ref{fig:CL_857_sensitivity} show the sensitivity of the power spectra to the model parameters at 217GHz and 857 GHz. In each panel, we increase or decrease a parameter by $2\sigma$. We note that the parameter $\sigma$ only affects the shot noise. Since shot noise dominates at larger $k$ and at higher frequency, the impacts of $\sigma$ is the strongest at large $k$ at 857 GHz.

\begin{table*}
\centering
\setlength{\tabcolsep}{0.5em}
\begin{tabular}{c|cccccccccccccc}
\hline
\rule[-2mm]{0mm}{6mm}&$\eta_0$&$\alpha_1$&$\alpha_2$&$\beta$&$\sigma$&$\delta$\\\hline
\rule[-2mm]{0mm}{6mm}$\eta_0$&1.00&-0.59&-0.88&-0.48&-0.07&0.99\\
\rule[-2mm]{0mm}{6mm}$\alpha_1$&-0.59&1.00&0.55&0.36&-0.47&-0.53\\
\rule[-2mm]{0mm}{6mm}$\alpha_2$&-0.88&0.55&1.00&0.50&0.26&-0.86\\
\rule[-2mm]{0mm}{6mm}$\beta$&-0.48&0.36&0.50&1.00&-0.05&-0.56\\
\rule[-2mm]{0mm}{6mm}$\sigma$&-0.07&-0.47&0.26&-0.05&1.00&-0.10\\
\rule[-2mm]{0mm}{6mm}$\delta$&0.99&-0.53&-0.86&-0.56&-0.10&1.00\\\hline
\end{tabular}
\caption{Correlation matrix for the model parameters.}
\label{tab:cor}
\end{table*}

\section{Complete figures of comparisons between observations and our model}\label{app:full}
\begin{figure*}
\centering
\includegraphics[width=2\columnwidth]{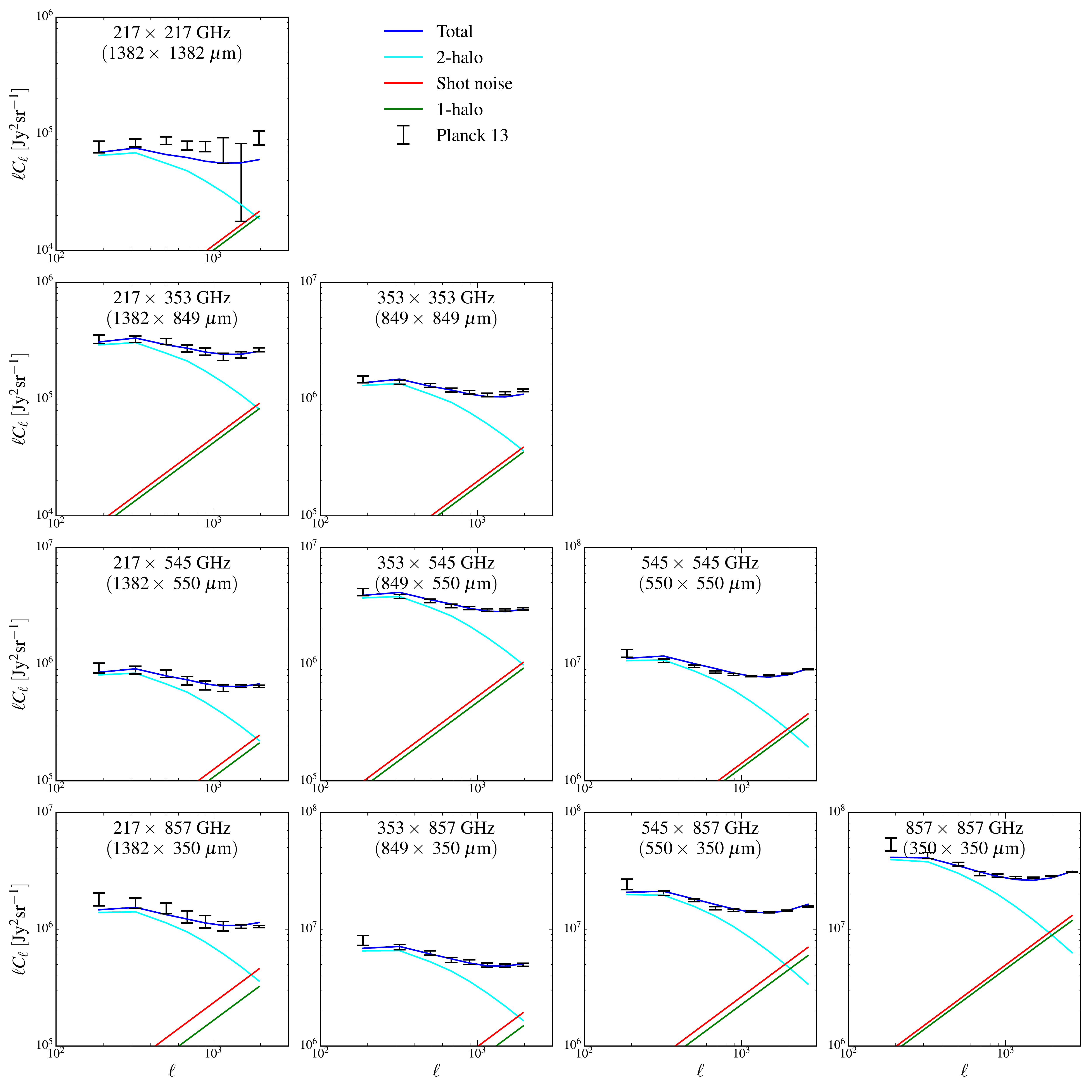}
\caption[{\em Planck} power spectra for CFIRB]{Our model fitting to the CFIRB power spectra from {\em Planck}.}
\label{fig:CL_full}
\end{figure*}

\begin{figure*}
\centering
\includegraphics[width=2\columnwidth]{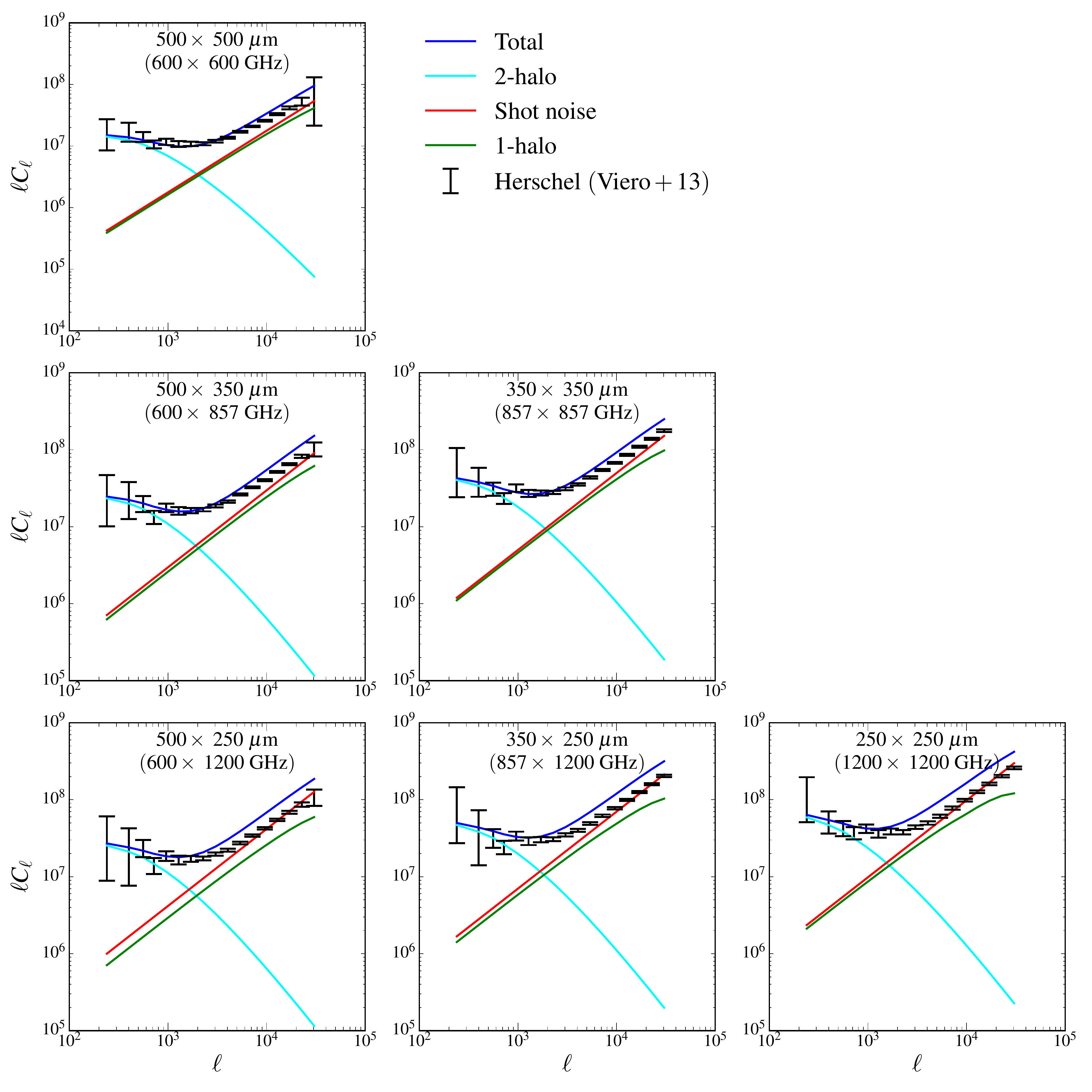}
\caption[{\em Herschel} power spectra for CFIRB]{CFIRB power spectra from {\em Herschel}--SPIRE (see Section~\ref{sec:CL_Herschel}) compared with our model prediction.}
\label{fig:CL_Herschel_full}
\end{figure*}

\begin{figure}
\centering
\includegraphics[width=1\columnwidth]{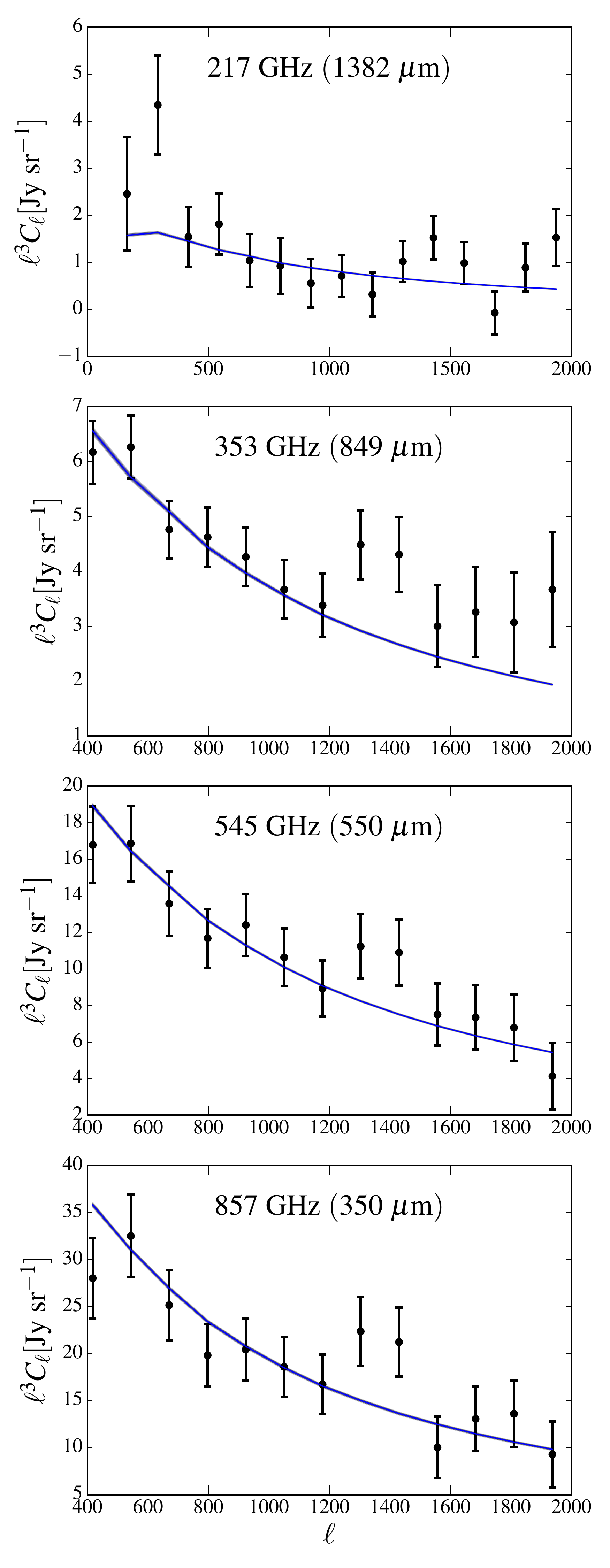}
\caption[]{Correlation between CFIRB and CMB lensing potential (see Section~\ref{sec:lensing}) compared with our model prediction.}
\label{fig:lensing_full}
\end{figure}

\begin{figure}
\centering
\includegraphics[width=1\columnwidth]{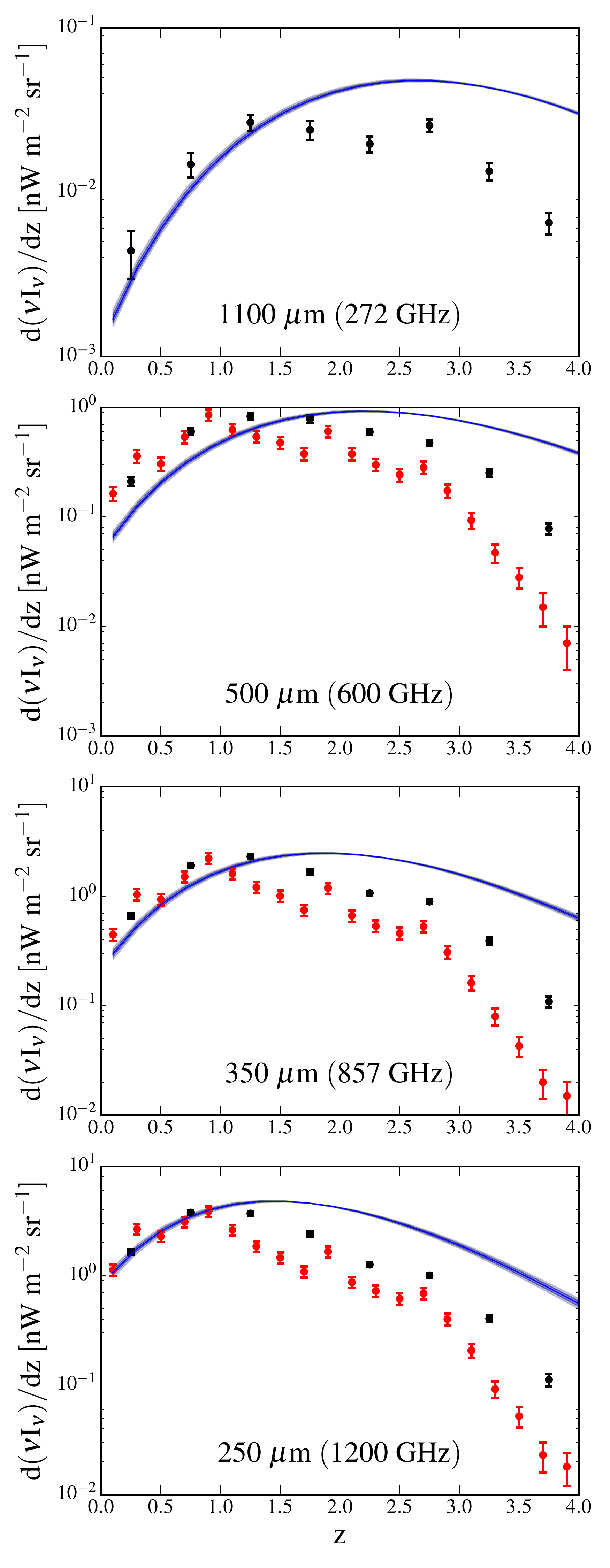}
\caption[]{Redshift distribution of CFIRB (see Section~\ref{sec:dInu}) compared with our model prediction.}
\label{fig:dInu_full}
\end{figure}

\begin{figure*}
\centering
\includegraphics[width=1.85\columnwidth]{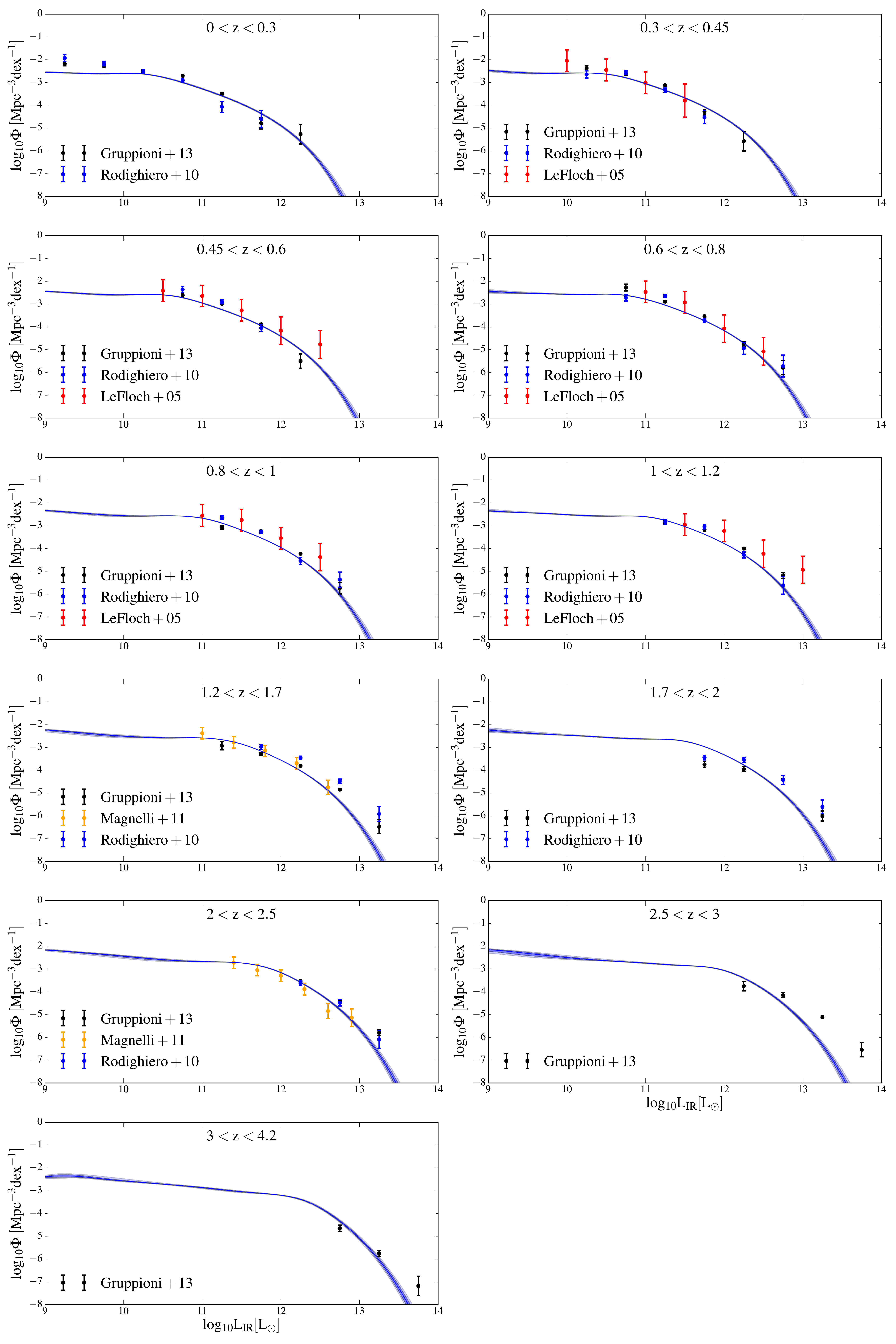}
\caption[]{Bolometric infrared luminosity functions (see Section~\ref{sec:LF}) compared with our model prediction.}
\label{fig:LF_full}
\end{figure*}

\begin{figure*}
\centering
\includegraphics[width=2\columnwidth]{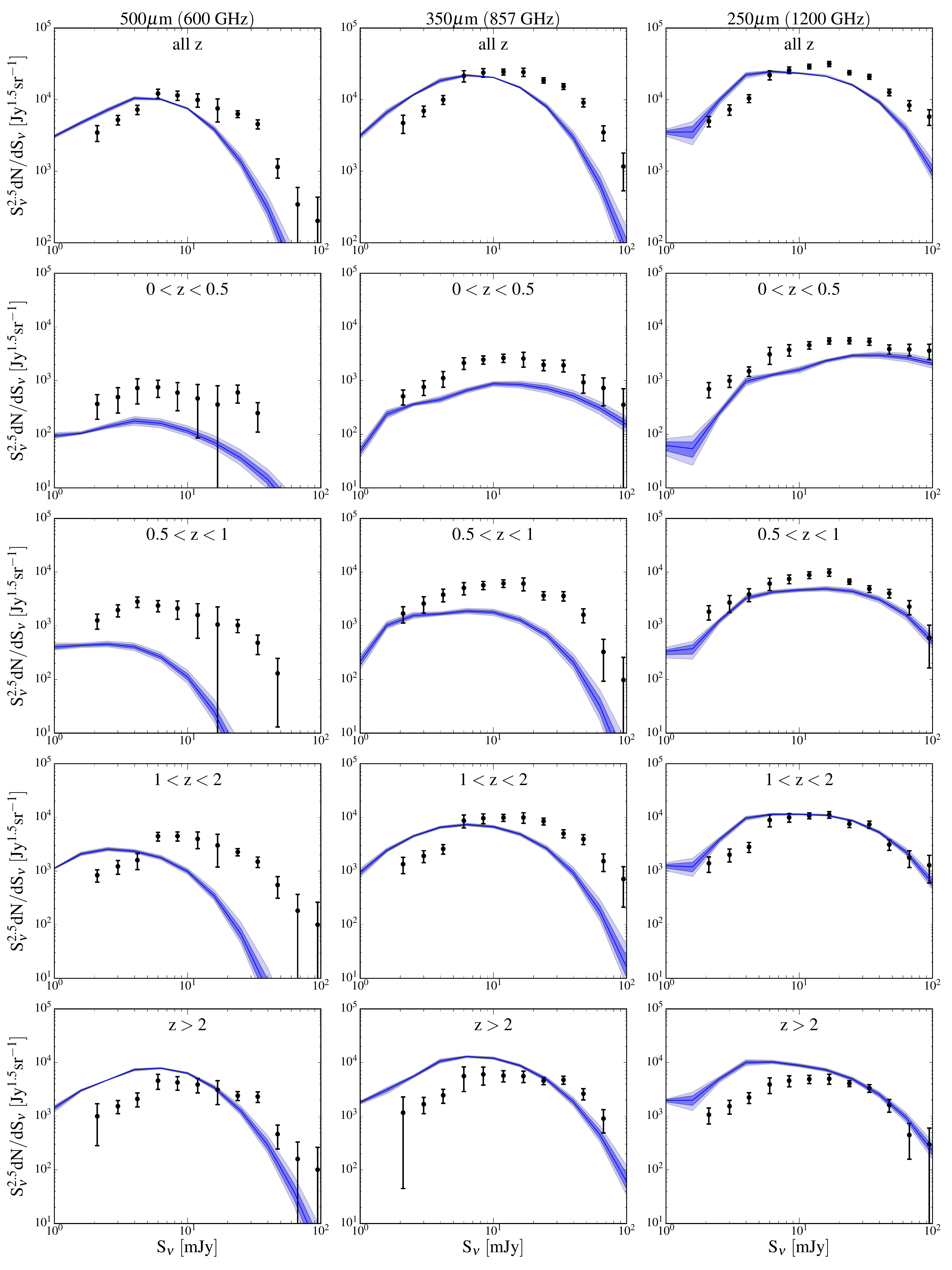}
\caption[]{Number counts data from \protect\citet[see Section~\ref{sec:NC}]{Bethermin12} compared with our model prediction.}
\label{fig:NC_full}
\end{figure*}

Most of the figures in the main text only show a single band or redshift slice for the purpose of demonstration. 
In this appendix, we show the full comparison between model predictions and observations we have conducted. 
\begin{itemize}
\item Fig.~\ref{fig:CL_full}: our fit to the {\em Planck} power spectra of CFIRB.
\item Fig.~\ref{fig:CL_Herschel_full}: our model prediction for the {\em Herschel} power spectra of CFIRB.
\item Fig.~\ref{fig:lensing_full}: our model prediction for the correlation between CFIRB and CMB lensing potential.
\item Fig.~\ref{fig:dInu_full}: our model prediction for the redshift distribution of CFIRB emission.
\item Fig.~\ref{fig:LF_full}: our model prediction for the bolometric IR luminosity functions.
\item Fig.~\ref{fig:NC_full}: our model prediction for the FIR flux density functions (number counts).
\end{itemize}

\bsp	
\label{lastpage}
\end{document}

%% file: mycommand.tex
\newcommand{\comment}[1]{}
\newcommand{\beq}{\begin{equation}}
\newcommand{\eeq}{\end{equation}}
\newcommand{\bal}{\begin{aligned}}
\newcommand{\eal}{\end{aligned}}
\newcommand{\beqa}{\begin{equation}\begin{aligned}}
\newcommand{\eeqa}{\end{aligned}\end{equation}}
\newcommand{\avg}[1]{\left\langle{#1}\right\rangle}

\newcommand{\Msun}{\rm M_\odot}
\newcommand{\Lsun}{\rm L_\odot}
\newcommand{\LUV}{L_{\rm UV}}
\newcommand{\LIR}{L_{\rm IR}}

\newcommand{\dMsf}{\dot{M}_{\rm sf}}
\newcommand{\tsf}{t_{\rm sf}}
\newcommand{\fga}{f_{\rm ga}}
\newcommand{\Mh}{M_{\rm h}}
\newcommand{\Mg}{M_{\rm g}}
\newcommand{\Ms}{M_{\rm s}}
\newcommand{\Mm}{M_{\rm m}}
\newcommand{\Md}{M_{\rm d}}
\newcommand{\dMh}{\dot{M}_{\rm h}}
\newcommand{\dMg}{\dot{M}_{\rm g}}
\newcommand{\dMs}{\dot{M}_{\rm s}}
\newcommand{\dMm}{\dot{M}_{\rm m}}
\newcommand{\Td}{T_{\rm d}}

\newcommand{\Mvir}{M_{\rm vir}}
\newcommand{\Vvir}{V_{\rm vir}}

\newcommand{\Lnuobs}{L_{(1+z)\nu}}

\newcommand{\lnL}{{\rm ln} L}
\newcommand{\lnSnu}{{\rm ln} S_\nu}
\newcommand{\lnM}{{\rm ln} M}

\newcommand{\dMa}{\dot{M}_{\rm a}}

\newcommand{\SFR}{{\rm SFR}}

\newcommand{\Mpk}{M_{\rm pk}}
\newcommand{\Mmin}{M_{\rm min}}

\newcommand{\etaCon}{1.08^{+0.07}_{-0.07}}

\newcommand{\etaPri}{[0.00,2.00]}
\newcommand{\alphaOneCon}{2.50^{+0.11}_{-0.07}}
\newcommand{\alphaOneBest}{2.50}
\newcommand{\alphaOnePri}{[0.00,3.00]}
\newcommand{\alphaTwoCon}{0.49^{+0.01}_{-0.01}}
\newcommand{\alphaTwoBest}{0.49}
\newcommand{\alphaTwoPri}{[0.00,3.00]}
\newcommand{\betaCon}{2.11^{+0.02}_{-0.02}}

\newcommand{\betaPri}{[1.50,2.50]}
\newcommand{\sigmaCon}{0.35^{+0.04}_{-0.03}}

\newcommand{\sigmaPri}{[0.20,2.00]}
\newcommand{\deltaCon}{1.13^{+0.06}_{-0.06}}

\newcommand{\deltaPri}{[-1.00,3.00]}
\newcommand{\chisqBest}{207.691}
\newcommand{\dof}{81}
\newcommand{\alphaOneBestTimesThree}{7.5}
\newcommand{\sigmaDex}{0.13}
\newcommand{\LIRnorm}{3.6\times10^{10}}
\newcommand{\factorfM}{2}
\newcommand{\Mdnorm}{14\times10^{6}}
\newcommand{\Tdnorm}{28}
\newcommand{\MpkFid}{12}